\newif\ifjrssb
\jrssbfalse

\ifjrssb
\documentclass{statsoc}
\usepackage[a4paper, top=1in, bottom=1in, left=1in, right=1in]{geometry}

\else
\documentclass[letterpaper,12pt]{article}
\usepackage{geometry}
\usepackage{authblk}
\fi

\usepackage{float}
\usepackage{subcaption}

\usepackage{amsmath}
\usepackage{commath}
\usepackage{amsthm}
\usepackage{amssymb}
\usepackage{setspace,mathrsfs}
\usepackage{algorithm}
\usepackage{algorithmic}
\usepackage{url,amsmath,amsfonts,amssymb}
\usepackage{graphicx}
\usepackage{listings,enumerate,color,relsize,multirow,rotating}
\usepackage{bbm}
\usepackage{natbib}
\usepackage{hyperref}
\hypersetup{colorlinks,citecolor=blue,urlcolor=blue,linkcolor=blue}
\usepackage{bm}
\usepackage{lipsum}
\usepackage[english]{babel}
\usepackage{enumitem}
\usepackage{footmisc}
\usepackage{makecell}
\usepackage{indentfirst}
\usepackage{subcaption}
\usepackage{multirow}
\usepackage{booktabs} 
\usepackage{diagbox}

\usepackage{geometry}

\definecolor{purple}{rgb}{0.84, 0.17, 0.89}
\definecolor{red2}{rgb}{0.7, 0, 0.1}
\definecolor{blue2}{rgb}{0.1,0.1,0.65}
\usepackage{color}

\usepackage{my_tex}

\def\Mix{{\rm Mix}}
\def\KL{{\rm KL}}
\def\mix{{\rm mix}}

\ifjrssb

\else
\title{Domain Adaptation Optimized for Robustness in Mixture Populations}
\author[1]{Keyao Zhan}
\author[1]{Xin Xiong}
\author[2]{Zijian Guo}
\author[1]{Tianxi Cai}
\author[3,4]{Molei Liu}

\affil[1]{Department of Biostatistics, Harvard T.H. Chan School of Public Health.}
\affil[2]{Department of Statistics, Rutgers University.}  
\affil[3]{Department of Biostatistics, Peking University Health Science Center}
\affil[4]{Beijing International Center for Mathematical Research, Peking University}
\fi

\begin{document}

\setstretch{1.05}

\maketitle
\date{}

\begin{abstract}

\noindent Integrative analysis of multi-institutional electronic health record (EHR) data can advance precision medicine by leveraging large, diverse datasets. However, generalizing findings to future populations is hindered by heterogeneity in patient demographics and clinical practices across institutions. While domain adaptation methods address data shifts, most assume target populations align with at least one source population, neglecting mixtures that combine sources influenced by factors like demographics. Additional challenges in EHR-based studies include unobserved outcomes and the need to explain population mixtures using broader clinical characteristics than those in standard risk models. To address these challenges under shifts in both covariate distributions and outcome models, we propose a novel framework: {\bf Domain Adaptation Optimized for Robustness in Mixture populations} (DORM). Leveraging partially labeled source data, DORM constructs an initial target outcome model under a joint source-mixture assumption. To enhance generalizability to future target populations that may deviate from the joint source-mixture approximation, DORM incorporates a group adversarial learning step to derive a final estimate, optimizing its worst-case performance within a convex uncertainty set built around the initial target model. In addition, this robust domain adaptation procedure is assisted by high-dimensional surrogates that enhance transferability in EHR studies. When a small set of gold-standard or noisy labels is available from the target population, a tuning strategy is implemented to refine the uncertainty set, mitigating conservativeness and further improving performance for the specific target population. Statistical convergence and predictive accuracy of our method are quantified through asymptotic studies. Simulation and real-world studies demonstrate the out-performance of our method over existing approaches.

\end{abstract}


\section{Introduction}

\label{sec:background}
Integrative analysis of multi-institutional biobank-linked electronic health record (EHR) data holds immense potential to advance precision medicine. By leveraging large and diverse datasets, such analyses can uncover novel insights and improve the performance of risk prediction models. However, ensuring the generalizability of these models to future target populations remains a significant challenge. This difficulty stems from the heterogeneity among patient populations, the presence of noise and bias in training datasets, and the potential underrepresentation of future target populations within existing datasets. For instance, the UK Biobank (UKB) comprises over 90\% participants of European descent, with less than 4\% of African or South Asian ancestry. This imbalance creates challenges in training genetic risk models optimized for underrepresented populations, particularly when the effects of genetic variants on disease risk vary across ancestries due to differences in genetic architecture, such as linkage disequilibrium (LD) structure \citep{west2017genomics,martin2019clinical}. Developing ancestry-specific genetic risk scores (GRS) has been proposed as a potential solution to improve prediction accuracy. However, there is a significant gap in the literature regarding the optimization of GRS for individuals with mixed ancestries, whose genetic profile may resemble a mixture of several. Moreover, patient populations vary inherently across biobanks, with each enrolling a distinct demographic composition. Effectively addressing these heterogeneities is crucial to ensuring the generalizability of prediction models across diverse future target populations \citep{de2021developing}.

Optimizing prediction performance for future target domain using datasets from multiple source domains has become a focal point of research in recent years. In the field of domain adaptation, a lot of knowledge transfer methods have been developed to leverage multiple source data sets and improve performance in a specified target population. For example, \cite{li2022transfer} and \cite{tian2023transfer} leveraged information from source data through sparse shrinkage on the target domain and introduced algorithm-free procedures to detect transferable ones from multiple sources. \cite{cai2022semi} and \cite{he2024transfusion} improved this so-called Trans-Lasso framework by addressing covariate shift and semi-supervised problems. \cite{gu2022robust} and \cite{lin2024profiled} extended Trans-Lasso to angle-based or profile transfer learning approaches, in which the target model can be approximated using a linear combination of the coefficients provided by the sources. In addition, \cite{tian2023learning} and \cite{li2023multi} both considered a multi-source setting with the coefficients of the source and target models embedded in a low-rank latent space. Although these approaches demonstrate solid utility and great potential in multi-source domain adaptation, most of these existing methods either assume that the outcome model for the target population is close to that of a subset of the source populations and/or require labeled data in the target population, and they are not efficiently leveraging the source-mixing target structure for efficient knowledge transfer.

When there are no labeled samples in the target domain, unsupervised domain adaptation (UDA) emerges to address domain shift between labeled source and unlabeled target domains, especially in the field of deep learning and computer science \citep{kouw2018introduction,liu2022deepunsuperviseddomainadaptation}. UDA methods bridge the gap between source and target domains typically using statistical moment matching (maximum mean discrepancy, \citep{long2018conditional}), feature-level adversarial learning \citep{ganin2016domainadversarialtrainingneuralnetworks}, self-training \citep{liu2021energy}, and so on.  However, most of these approaches generally do not fully exploit the knowledge available from multiple source datasets, nor do they simultaneously address both covariate shift and posterior drift. In addition, these methods do not account for the potential mixture structures within the target domain.

Recent work on data integration and federated learning with a mixture of heterogeneous source datasets is also relevant to our setup in the sense that the local sites are assumed to be the mixture of some latent subgroups with unknown probabilities. For this problem, \cite{marfoq2021federated} proposed an EM algorithm to achieve effective data integration. \cite{tian2022unsupervised} and \cite{wu2023personalized} tackled the specific Gaussian mixture problem in a similar context.  Nevertheless, the fundamental goal of this track is to pursue an integrative model working well for the local source clients in an average sense, which is essentially different from our task with a potential future mixture target domain. In addition, robustness to deviation from such mixture or latent subgroup assumptions is a crucial but understudied question in these existing literature.

Existing domain adaptation or data integration methods often struggle to ensure generalizability across a diverse range of future target populations. To address this, group distributionally robust optimization (DRO) methods have been developed, aiming to enhance generalizability to a broader collection of future populations. These methods typically define an uncertainty set as a mixture of observed source populations and optimize a model to perform robustly by minimizing its worst-case performance within this set. For instance, \cite{meinshausen2015maximin} introduced a maximin framework to maximize the minimum reduced variance of a linear model across multiple data sources. Building on this, \cite{guo2022statistical} developed a resampling approach for non-normal and high-dimensional inference on maximin effects, while \cite{wang2023distributionally} extended this framework to general machine learning. In machine learning, \cite{sagawa2019distributionally} addressed poor generalization in deep neural networks by minimizing worst-case loss over predefined groups with added regularization, \cite{ijcai2023p162} used density ratio for learning calibrated uncertainties
under domain shifts, and \cite{ghosal2023distributionally} explored probabilistic group DRO with subject-specific group probabilities. These methods often achieve robustness at the cost of being overly conservative for certain target populations due to the large model space the uncertainty set contains.

When prior knowledge about future target populations is available, it is beneficial to incorporate this knowledge to reduce the conservativeness of standard DRO methods. For instance, when a small subset of labeled observations is available from a target population, \cite{xiong2023distributionally} proposed a distributionally robust domain adaptation approach that utilizes these observations to guide the group DRO framework, ensuring improved predictive performance on future populations similar to the observed target. \cite{mo2024minimax} introduced a group minimax regret framework, which replaces the baseline null model in maximin regression with the empirical risk minimizer. In addition, \cite{wang2025knowledge} constructs less conservative ambiguity sets for DRO by controlling the transportation along directions informed by the source knowledge, which protects against information loss. However, these existing methods still require a sufficiently large number of observed labeled samples from the target to effectively train a target-only model, raising a challenge in the typical absence or scarcity of labels of mixed ancestries in Biobank data.

\subsection{Our results and contributions}

Motivated by deriving generalizable GRS for admixture populations, we consider an understudied generalizable multi-source domain adaptation setting, assuming the joint distribution of the covariates and outcomes on the target domain can be well-approximated by a mixture of the source distributions with unknown weights. The ideal joint-mixing assumption is given by equation (\ref{eq:ideal joint}) while we also accommodate potential violations on this as demonstrated in Figure \ref{fig: data}. To handle this problem, we propose a novel framework as \textbf{D}omain adaptation \textbf{O}ptimized for \textbf{R}obustness in \textbf{M}ixture populations (DORM).

DORM first leverages general machine learning (ML) methods to approximate the covariate distribution on the target domain with a mixture of the sources. A surrogate-assisted and model-assisted construction is used to adjust for covariate shift between the sources and target while maintaining robustness to errors of ML estimation. Then, DORM incorporates group adversarial learning in order to maintain distributional robustness to moderate violation of the source-mixing assumption (\ref{eq:ideal joint}) imposed on the target population. Our framework also allows flexible specification and tuning on the degree of uncertainty in this DRO procedure to ensure good performance against the violation of our assumption as well as avoid over-conservativeness. Through theoretical analysis, DORM is shown to achieve desirable convergence to its population value, as well as good predictive performance on the target when the joint mixing assumption tends to hold. In terms of prediction and out-of-distribution generalization, DORM outperforms existing domain adaptation methods in our simulation and real-world studies. Regarding comprehensive literature, the novelty and main contributions of our work can be summarized as follows.

First, DORM fills the gap of existing multi-source domain adaptation tools in addressing the source-mixture type of targets as introduced in Section \ref{sec:background}. Compared to existing approaches like \cite{li2022transfer} and \cite{xiong2023distributionally} relying sufficient labeled target sample for high-dimensional regression, a main advantage of leveraging the source-mixing structure in DORM is to realize knowledge transfer in a more challenging setup in scarcity or even absence of labeled target sample, but still under the outcome model shift between the sources and target. Although such challenges are frequently encountered in various application scenarios such as EHR and biobank studies, effective and robust knowledge transfer has not been easily achieved in this scenario due to the lack of methods leveraging the covariate similarity to infer the outcome model on the target population. This gap is mitigated by our proposed scheme mixing the joint distribution of $(X,Y)$ from the sources but not directly assembling the regression coefficients like \cite{xiong2023distributionally} or \cite{lin2024profiled}.

Second, DORM provides a flexible framework to realize a better trade-off between the predictive performance on some pre-defined target distribution and the generalizability out of this distribution. Most existing DRO methods fail to predict well on testing data since their uncertainty sets of adversarial distributions are specified to be overly general and large, which usually results in over-conservativeness, e.g., the whole simplex of the source models as used in \cite{meinshausen2015maximin}. In DORM, we address this problem by deriving a guidance for the uncertainty set based on the joint-mixing assumption for the target distribution, which is in more favor of the predictive performance on target compared to existing DRO methods. Meanwhile, our adversarial learning part ensures robustness to the possible deviation of the actual target from such assumptions, as well as generalizability to some future distributions moderately different from the specific target domain. Such generalization has not been realized by recent work focusing on some specific target \citep[e.g.]{li2022transfer,lin2024profiled}. In this context, our method allows flexible choice on the degree of uncertainty by leveraging side information and according to the real needs.

Finally, although surrogate-assisted approaches have already raised great research interest in fields such as semi-supervised learning \citep[e.g.]{hou2023surrogate,xia2024prediction}, there is still a lack of work utilizing surrogates to realize multi-source domain adaptation targeting mixture populations or group adversarial learning. Our work fills this gap by effectively leveraging the surrogate features to identify source-mixing approximation for the target domain. Noting that distributional shift of the high-dimensional surrogates across the source and target domains could cause excessive bias, we address this technical challenge through a double machine learning (DML) framework shown to be insensitive to the errors of ML models for covariate shift correction.

\section{DORM: Definition and Identification}

Suppose that we have access to $L$ source datasets $\{X^{(l)}, Y^{(l)}\}_{1 \leqslant l \leqslant L}$ collected from different sub-populations. For $1 \leqslant l \leqslant L$, assume that $\{X_i^{(l)}, Y_i^{(l)}\}_{1 \leqslant i \leqslant n_l}$ are generated following
\[
    X_i^{(l)} \stackrel{i . i . d}{\sim} \mathbb{P}_X^{(l)}, \quad Y_i^{(l)} \mid X_i^{(l)} \stackrel{i . i . d}{\sim} \mathbb{P}_{Y \mid X}^{(l)} \quad \text { for } \quad i \in [n_l],
\]
where $\mathbb{P}_X^{(l)}$ denotes the distribution function of $X_i^{(l)} \in \mathbb{R}^p$, $\mathbb{P}_{Y \mid X}^{(l)}$ denotes the conditional distribution of the outcome $Y_i^{(l)} \in \mathbb{R}$ given $X_i^{(l)}$, and $[n]=\{1, \ldots, n\}$ for any positive integer $n$. To accommodate semi-supervised settings common in EHR, where observing $Y$ often requires manual annotation, we allow each source to additionally observe unlabeled samples with only covariates $X\ul_i$ observed for $n_l < i \leqslant N_l$. Suppose that the covariates $X = (A\trans,W\trans)\trans$, where $A_{q\times 1}$ denotes a vector of predictors for $Y$ with its first element being constant $1$ and $W_{(p-q)\times 1}$ denotes some auxiliary covariates that are informative for $Y$ but not included in the outcome model of our interest. In our primarily interested setup, $X$ can be high-dimensional and $A$ can be low- or high-dimensional. In summary, for each source $l \in [L]$, we observe the data $\Dsc_l =\big\{X\ul_i= (A_i\ul,W_i\ul)_{i\in [N_l]}, (Y_i\ul)_{i\in [n_l]}\big\}$.

For a potential target population, assume the underlying covariates and outcome, denoted as $\big\{X_i\uz=(A_i\uz,W_i\uz),Y_i\uz\big\}$, follow
\[
    X_i\uz \stackrel{i . i . d}{\sim} \mathbb{P}\xz, \quad Y_i\uz \mid X_i\uz \stackrel{i . i . d}{\sim} \mathbb{P}_{Y \mid X}\uz \quad \text { for } \quad i \in [N_0].
\]
Our objective is to develop a predictive and generalizable linear risk model for $Y \sim A$ in the target domain by using multiple source datasets, as explained further in the following Remark \ref{rem:1}. Beyond achieving strong predictive performance for a specific target population, we aim to ensure distributional robustness, enabling the model to perform reliably across a range of potential target populations. This is challenging due to two key types of distributional shifts: (i) covariate shift: $\mathbb{P}\xz$ and $\{\mathbb{P}_X^{(l)}\}_{l \in[L]}$ can be all different; (ii) posterior drift: $\mathbb{P}_{Y \mid X}\uz$ and $\{\mathbb{P}_{Y \mid X}^{(l)}\}_{l \in[L]}$ can be all different. For now, we focus on the unsupervised domain adaptation regime where we only have access to covariate observations $X_i\uz$ but no outcome observations $Y_i\uz$ on the target. Specifically, we only observe the data $\Dsc_0 = \{X_i\uz=(A_i\uz,W_i\uz)_{i\in [N_0]}\}$. This scenario is both challenging and common in real-world applications such as EHR and biobank studies, where obtaining outcome labels is costly.

\begin{remark}
Auxiliary or surrogate features $W$ have been frequently considered in related literature \citep[e.g.]{liu2023augmented}. Taking EHR-linked genetic risk studies as an example, $Y$ is the primary phenotype obtained by manual chart review, $A$ is taken as genetic markers and $W$ includes EHR proxies of $Y$ such as relevant diagnostic codes and laboratory test results. Since the scientific goal is to use genes to predict disease risk, our interest is in modeling $Y\sim A$ rather than $Y\sim A,W$. Nevertheless, $W$ can still serve as nuisance features since it is not only informative to $Y$ but also a characteristic of the distributional shift between the sources and the target. The inclusion of $W$ can help characterize latent confounders or mediators; thus, while the conditional distribution $Y\mid A$ may differ substantially across populations, conditioning further on $W$ can often reduce these discrepancies, making $Y\mid A,W$ more stable and transferable.
\label{rem:1}
\end{remark}

\begin{figure}[!htb]
    \centering
    \includegraphics[width=.6\linewidth, page=2]{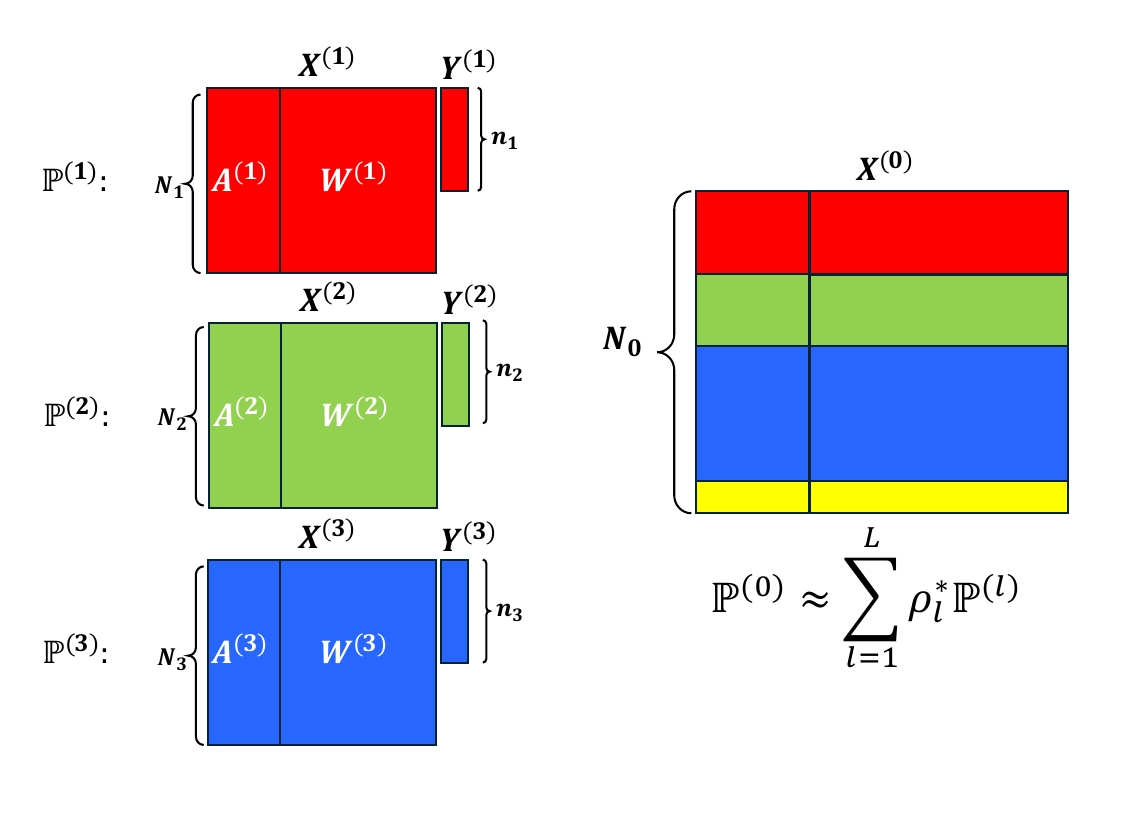}
    \caption{Data structure in the setting of DORM.  For each source domain $l \in [L]$, we can observe data $\Dsc_l =\{X\ul_i= (A_i\ul,W_i\ul)_{i\in [N_l]}, (Y_i\ul)_{i\in [n_l]}\}$, while on the target domain we can only observe $\{X_i\uz=(A_i\uz,W_i\uz)_{i\in[N_0]}\}$. The target joint distribution is approximately the mixture of source joint distributions, where yellow square in the target population is the density not belonging to any source population, representing the violation extent to our ``joint-mixing'' assumption. }
    \label{fig: data}
\end{figure}

To ensure distributional robustness while preserving accuracy for future target populations similar to the observed target, we introduce a {\em joint-mixing} working assumption. This assumption enables us to derive an initial estimate of the target distribution $\mathbb{P}_{Y \mid X}\uz$, providing a foundation for robust model adaptation. Next, we develop the DORM estimator, designed to optimize the worst-case performance across a set of potential target distributions. This set is defined within the convex hull spanned by the estimated target distribution and the observed source distributions, ensuring robustness against distributional shifts. Figure \ref{fig: data} illustrates the data structure and the approximate joint-mixing assumption to be formally introduced in the next sections.

\subsection{Key joint-mixing distribution assumption}
Under the {\em joint-mixing} assumption, the joint distribution $(\mathbb{P}\xz,\mathbb{P}_{Y \mid X}\uz)$ on the target domain is close to a mixture of the source distributions $(\mathbb{P}^{(l)}_X,\mathbb{P}_{Y \mid X}^{(l)})$ for $l\in[L]$. The ideal version of this joint-mixing assumption can be expressed as
\begin{equation}
    (X^{(0)},Y^{(0)}) \sim \mathbb{P}\uz:= \mathbb{P}\xz\mathbb{P}_{Y \mid X}\uz =\suml \rho_l^* \mathbb{P}\ul:= \suml \rho_l^* \mathbb{P}^{(l)}_{X}\mathbb{P}^{(l)}_{Y\mid X},
\label{eq:ideal joint}
\end{equation}  
where $\rho^*=(\rho_1^*,\ldots,\rho_L^*)$ is the set of (prior) mixing probabilities belonging to the $L$-dimensional simplex 
$
\Delta_L = \{\rho \in \mathbb{R}_L \mid \suml \rho_l = 1, \rho_l \geqslant 0 , \forall l \in [L]\}.
$
Note that assumption (\ref{eq:ideal joint}) also implies a mixture form of the marginal distribution of $X$:
\begin{equation}
    \mathbb{P}\xz=\sum_{l=1}^L \rho_l^* \mathbb{P}_X^{(l)}.
\label{eq:xmix}
\end{equation}
But for identifying the conditional distribution for target domain $\PP_{Y|X}\uz$, it's not a simple linear combination weighted by $\rho^*$. Let $d\PP$ denote the probability density or mass function of any distribution function $\PP$. For any $\rho\in \Delta_L$, denote the mixture density function as
\[
\Mix(x;\rho)=\sum_{l=1}^L\rho_l d\PP^{(l)}_X(x).
\]
To sample $X$ following $\Mix(x;\rho)$, one can first generate a latent variable $S\in [L]$ with $\PP(S = l)=\rho_l$ indicating the original source, and then sample $X \mid S = l \sim \mathbb{P}\xl$. Then when the ideal joint-mixing assumption (\ref{eq:ideal joint}) strictly holds, we have $X_i^{(0)}\sim \Mix(x;\rho^*)$. As will be shown in the Proposition \ref{propo:conmodel} below, under assumption (\ref{eq:ideal joint}), $\mathbb{P}_{Y \mid X}\uz$ is identifiable from $\Dsc_l$'s and $\Dsc_0$ even without any observations of $Y^{(0)}$ on the target domain. This will be particularly useful for the ``joint-mixing" type of target populations introduced in Section \ref{sec:background}.

\begin{proposition}
    \label{propo:conmodel}
    If the ideal joint-mixing assumption (\ref{eq:ideal joint}) strictly holds, then
    \begin{equation}
    \mathbb{P}_{Y \mid X}\uz = \sum_{l=1}^L \eta_l(X;\rho^*) \mathbb{P}_{Y \mid X}\ul
    \label{eq:y|xmix}
    \end{equation}
    where the posterior probability $\eta_l(X;\rho^*)$ is defined as
    \begin{equation*}
        \eta_l (X;\rho^*):= { \mathbb{P}} (S=l\mid X)=\frac{\rho^*_l d\PP^{(l)}_X(X)}{\sum_{k=1}^L\rho^*_k d\PP^{(k)}_X(X)}.
    \end{equation*}
\end{proposition}
When it comes to identifying $\rho$ from our data, without observations of $Y$ on the target, it is natural to find such $\rho$ through the best mixture approximation of covariates $X$
\begin{equation}
  \bar\rho= \underset{\rho \in \Delta_L}{\arg\min}\ \KL\left( d\PP_X^{(0)}\mid \Mix(x;\rho) \right)=\underset{\rho \in \Delta_L}{\arg\max}\ \EE_{\PP_X^{(0)}}\log\{\Mix(X;\rho)\},
\label{eq:kldiv}
\end{equation}
where $\KL(\cdot\mid\cdot)$ represents the Kullback–Leibler (KL) divergence. Under certain assumptions, (\ref{eq:kldiv}) is a strongly convex problem that typically has a unique solution regardless of the condition in (\ref{eq:ideal joint}); and when (\ref{eq:ideal joint}) holds, we have $\bar\rho=\rho^*$ and $\Mix(x;\bar\rho)=\PP_X^{(0)}$; see more details in the theoretical justification in Section \ref{sec:theory}.

Under the joint-mixing assumption, we may obtain an estimate of the risk model $\beta$ via maximizing the reward function $R_\PP (\beta)$. For a generic model $\beta\in\mathbb{R}^q$, we define the reward function as $R_\PP (\beta) = \mathbb{E}_{(X, Y) \sim \PP} \left[Y^2-(Y-{ A\trans \beta})^2 \right]$ which measures the variance of $Y$ explained by $A\trans\beta$ when $(X,Y)$ is generated from distribution $\PP$. Then under the joint-mixing assumption, we have

\begin{equation}
\bar{\beta}_{\mix}=\underset{\beta\in\mathbb{R}^q}{\arg\max}\ \mathbb{E}_{(X, Y) \sim (\mathbb{P}\xz, \sum_{l=1}^L \eta_l(X;\bar\rho) \mathbb{P}_{Y \mid X}\ul)} \left[Y^2-(Y-{ A\trans \beta})^2\right],
\label{eq:least:sq:mix}
\end{equation}
with $\bar{\rho}$ enabling proper approximation of the target domain with source mixture distribution derived from (\ref{eq:kldiv}). Note that $\bar{\beta}_{\mix}$ is identifiable in our data setup without any observations of $Y_i^{(0)}$. We will discuss the reward function choice in the following Remark \ref{remark:reward}.

\subsection{DORM definition}
To attain distributional robustness, DORM derives a group distributional robust estimator by optimizing the worst case performance over a source-mixing uncertainty set for target domain distribution:
\begin{equation}
    \mathcal{M}\left(\smax\right)=\left\{ \left(\PP\xz,\mathbb{T}_{Y \mid X}(y;s,\delta)\right):\delta \in \Delta_L,~s \in [0,\smax]\right\} ,
\label{eq:uncertainty}
\end{equation}
where the potential target distributions are defined as
\begin{equation}
    \mathbb{T}_{Y \mid X}(y;s,\delta)=(1-s)\left[ \sum_{l=1}^L \eta_l(X;\bar\rho) \mathbb{P}_{Y \mid X}^{(l)}(y) \right]+ s \sum_{l=1}^L \delta_l \mathbb{P}_{Y \mid X}^{(l)}(y).
    \label{eq:Y|Xdist}
\end{equation}
This is essentially a mixture of $L+1$ different conditional distributions, where the first term in $\mathbb{T}_{Y \mid X}(y;s,\delta)$ corresponds to the conditional mixture distribution of $Y$ implied by (\ref{eq:ideal joint}) and (\ref{eq:y|xmix}), representing our best approximation of target $\PP\uz_{Y|X}$ in the ideal joint-mixing scenario. By assigning a large proportion \(1-s\) to this mixture, we capture the intuition that (\ref{eq:ideal joint}) is our primary working assumption. The second term is a marginal mixture of source conditional distributions weighted by \(s\), where \(\delta=(\delta_1,\ldots,\delta_L)\) resides in the probability simplex to encode adversarial weights across sources \(l=1,\ldots,L\). Conceptually, this term allows part of the target distribution to deviate arbitrarily from our best guess, ensuring robustness against potential misspecifications and enhancing generalizability for various target domains. Here, $\smax$ is a hyper-parameter controlling the upper range of $s$, i.e., our trust in assumption (\ref{eq:ideal joint}), which will be discussed in more details later. 

Based on (\ref{eq:uncertainty}), we define the population level DORM regression coefficients as
\begin{equation}
\barbetare(\smax) :=\underset{\beta \in \mathbb{R}^q}{\arg \max } \min _{\PP \in \uncer} R_{\PP}(\beta),
\label{eq:droY}
\end{equation}
where the reward function $R_{\PP}(\beta):=\mathbb{E}_{(X, Y) \sim \PP}\left[Y^2-(Y-{ A\trans \beta})^2\right]$ measures the explained variance of $Y\sim A\trans \beta$ compared to the null model on $\PP$. In a similar spirit to group DRO, (\ref{eq:droY}) aims at optimizing the worst-case predictive performance on the uncertainty set $\uncer$. It can be viewed as a two-sided game that one agent searches over $\uncer$ to find the most adverse target distribution $\PP$ for $\beta$ and another agent updates $\beta$ to maximize the reward for such an unfavorable target population. Different from \cite{li2022transfer} and \cite{xiong2023distributionally}, solving (\ref{eq:droY}) does not rely on any observations of $Y^{(0)}$ on the target domain.

\begin{remark} \label{remark:reward}
Choice on the reward function $R_{\mathbb{P}}(\beta)$ is diverse in various studies. \cite{agarwal22b} and
\cite{mo2024minimax} frame their reward function as a regret, measuring the excess risk of the model compared to the optimal model for the distribution $\PP$. \cite{xiong2023distributionally} compared the mean squared loss of the model $\beta$ to that of the initial estimator $\beta_{\operatorname{init}}$ constructed using labeled target sample. \cite{wang2023distributionally} provided an evaluation of different reward function choices in the context of group adversarial learning, asserting that the reward function $R_{\mathbb{P}}(\beta)$ employed in our DORM framework is both independent of noise levels for source domains and computationally feasible when incorporating general convex prior mixture information.
\end{remark}

Importantly, the hyper-parameter $\smax$ in (\ref{eq:droY}) encodes a trade-off on the degree of adversary. The smaller $\smax$ gets, the smaller $\Msc (\smax)$ is, and the closer $\barbetare(\smax)$ will be to $\bar{\beta}_{\mix}$ that works under the ideal case (\ref{eq:ideal joint}). When $\smax=0$, we have $\barbetare(\smax)=\bar{\beta}_{\mix}$ without robustness to any adversarial distributions departing from (\ref{eq:ideal joint}). On the other hand, a larger $\smax$ results in a larger uncertainty set more likely to contain the actual target $\mathbb{P}\uz$ and produces a more robust model. When $\smax = 1$, $\barbetare(\smax)$ will approach the covariate-shift-adjusted maximin model introduced in \cite{guo2022statistical}. We will discuss potential tuning strategies of $\smax$ in Section \ref{sec:tuning} and study the impact of $\smax$ in both theoretical and numerical studies.  

\begin{figure}
    \centering
    \includegraphics[width=1\linewidth, page=1]{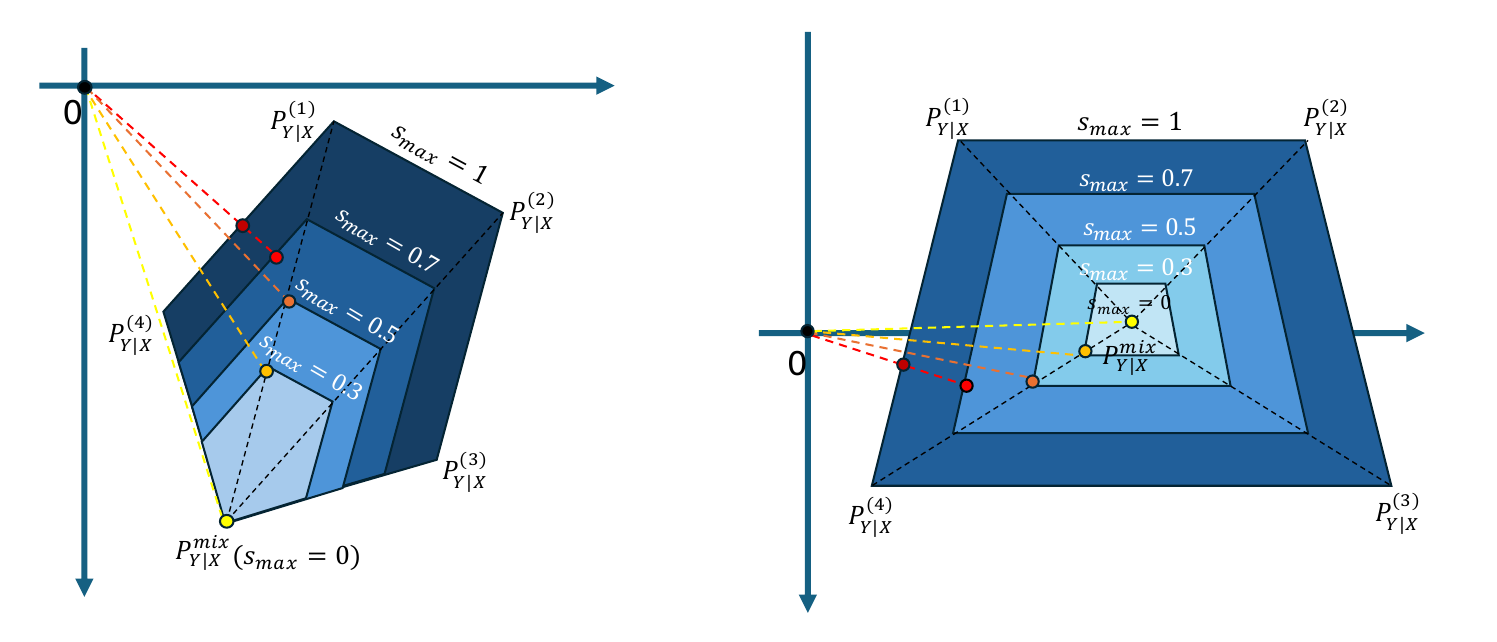}
    \caption{Geometric illustration of DORM uncertainty sets $\uncer$ (shaded area) and the corresponding solution $\barbetare(\smax)$ (colored dots) of (\ref{eq:droY}) with five different $\smax$ values. The left subplot shows the case when $\bar{\beta}_{\mix}$ (which represents the conditional mixture model) lies outside the convex hull spanned by the four sources, while the right represents $\bar{\beta}_{\mix}$ inside the source hull.}
    \label{fig: flowchart 1}
\end{figure}

\begin{remark}
     It is noteworthy that $\bar\beta_{\mix}$ is not a marginal mixture of each source model, because the $Y\mid X$ distribution $\sum_{l=1}^L \eta_l(X;\rho) \mathbb{P}_{Y \mid X}\ul$ in (\ref{eq:least:sq:mix}) is weighted by posterior weights which depend on $X$ rather than constant weights. Figure \ref{fig: flowchart 1} presents two different scenarios for $\bar\beta_\mix$. In the first scenario where $\bar\beta_\mix$ is outside the convex hull of sources, the uncertainty set for $Y\mid X$ conditional model in (\ref{eq:Y|Xdist}) is a convex combination of each source conditional and the mixture conditional. In the second scenario when $\bar\beta_{\mix}$ is already in the convex hull of source models, the uncertainty set for $Y\mid X$ is essentially a convex combination of source conditional models.
\end{remark}

\subsection{DORM identification}

For identification of $\barbetare(\smax)$, we derive its explicit form in the following theorem.
\begin{theorem}
The solution to (\ref{eq:droY}) can be identified from:
\begin{equation}
        \barbetare(\smax) = \sum_{l=1}^{L+1} \gamma^*_l \bar{\beta}_l \quad \text{with} \quad \gamma^* = \underset{\gamma \in \mathcal{S}(\smax)}{\arg \min } \gamma\trans \Gamma \gamma,
    \label{eq:idenPopu}
\end{equation}
where $\mathcal{S}(\smax) = \left\{\gamma \mid \gamma \in \Delta_{L+1}, \gamma_{L+1} \in [1-\smax,1] \right\}$, $\bar{\beta}_{L+1} :=\bar{\beta}_{\mix}$ is as defined in (\ref{eq:least:sq:mix}),
\begin{equation}
    \bar{\beta}_l=  \underset{\beta \in \mathbb{R}^q}{\arg\min}\ \EE_{(X,Y) \sim (\PP_X^{(0)},\PP_{Y \mid X}\ul)}\left(Y-A\trans \beta\right)^2,
    \label{eq:Popuql}
\end{equation}
the matrix $\Gamma=(\Gamma_{l,k})_{l,k \in [L+1]}$ with $\Gamma_{l,k}=\bar{\beta}_l\trans \bar\Sigma\uz \bar{\beta}_k$, and $\bar\Sigma\uz=\EE_0 AA\trans$ with $\EE_0:=\EE_{\PP_X^{(0)}}$. 
\label{thm:PopuIdenti}
\end{theorem}

By Theorem \ref{thm:PopuIdenti}, \(\barbetare(\smax)\) is revealed to be a convex combination of the regression coefficients \(\bar{\beta}_l\) (\(l \in [L]\)) and the mixture-based coefficient \(\bar{\beta}_{\mix}\). Each \(\bar{\beta}_l\) is obtained by a population-level least squares fit of \(Y\sim A\) under the hypothetical conditional distribution $(\PP_X^{(0)},\PP_{Y \mid X}\ul)$. Meanwhile, \(\bar{\beta}_{\mix}\) corresponds to the least squares solution under the mixture $(\PP_X^{(0)}, \sum_{l=1}^L \eta_l(X;\bar\rho) \mathbb{P}_{Y \mid X}\ul)$, which is also given in (\ref{eq:least:sq:mix}). Note that our framework permits misspecification of the linear model \(Y \sim A\) on each source domain, a flexibility accommodated by the semiparametric approach to be introduced in Section~\ref{sec:method:est}, which is different from the maximin regression formulations in \cite{meinshausen2015maximin} and \cite{guo2022statistical}.

The vector of weights \(\gamma^*\) in \eqref{eq:idenPopu} is found by a constrained quadratic optimization that balances these candidate coefficient vectors through the matrix \(\Gamma\). Notably, \(\gamma_{L+1}^*\ge 1 - \smax\) enforces that the final estimator always assigns a non-trivial weight to the mixture-based coefficient \(\bar{\beta}_{\mix}\), reflecting trust in the mixture model, which represents a principled way to hedge against potential but limited deviation from the idealized mixture assumption. This construction ultimately yields a robust and interpretable estimator that interpolates between the mixture-based regression and adversarially reweighted alternatives, thereby enhancing predictive reliability in heterogeneous or partially misspecified settings. Figure \ref{fig: flowchart 1} illustrates the uncertainty set in (\ref{eq:droY}) and demonstrates the solutions given by Theorem \ref{thm:PopuIdenti}.

\section{Estimation procedures of DORM}\label{sec:method:est}
\subsection{Doubly robust construction}

Let $\bar\eta_l(x)=\eta_l(x;\bar\rho)$, $\cpzh = N_0\inv \sum_{i=1}^{N_0}$ be the empirical mean operator on the target data $\Dsc_0$ and $\cplh = n_l\inv \sum_{i=1}^{n_l}$ denote the mean on the labeled sample from source $l$. For $u=(u_1,\ldots,u_d)\trans \in \mathbb{R}^d$, we define its $\ell_r$ norm as $\|u\|_r=\big(\sum_{j=1}^d|u_j|^r\big)^{1 / r}$ for any $r > 0$. Following Theorem \ref{thm:PopuIdenti}, the key step in finite sample implementation is to estimate coefficients $\bar{\beta}_l$ for $l\in[L+1]$ where $\bar{\beta}_{L+1}=\bar{\beta}_{\mix}$. Here we use $L+1$  for mathematical convenience in expressing vectors and matrices.
According to their definition, we have
\begin{equation}
    \bar{\beta}_l =(\bar\Sigma\uz)^{-1} \EE_0 \bar{m}_l(X) A,~l\in[L]; \quad\bar{\beta}_{\mix} =(\bar\Sigma\uz)^{-1} \suml \EE_0 \bar\eta_l(X)\bar{m}_l(X) A: = \bar{\beta}_{L+1},
    \label{eq:betabar}
\end{equation}
where $\bar{m}_l(x) = \EE_{\PP_{Y \mid X}\ul}(Y\mid X = x)$ is the conditional mean of $Y\mid X$ on source $l$. Suppose that we already obtain the estimators $\mlh$ and $\elh$ for the nuisance functions $\bar{m}_l$ and $\bar{\eta}_l$ using general learning methods like parametric regression and modern machine learning tools, the specific procedure of which will be introduced in Section \ref{sec:method:est:nuis}. Then for low-dimensional regression of $Y\sim A$, we can derive the plug-in estimators as:
\begin{equation}
    \widetilde{\beta}_l = (\szh)\inv \cpzh  \mlh(X_i\uz) A_i\uz, ~l \in [L]; \quad\widetilde{\beta}_{\mix} = (\szh)\inv  \suml \cpzh\elh(X_i\uz)  \mlh(X_i\uz) A_i\uz :=\widetilde{\beta}_{L+1},
    \label{eq:btil}
\end{equation}
where $\szh = \cpzh A\uz_i (A\uz_i)\trans$. Nevertheless, such $\widetilde{\beta}_l$'s may suffer from large bias due to the potential misspecification or excessive estimation errors in $\mlh$, especially when it is obtained with a relatively small labeled sample with the size $n_l$. Inspired by the doubly robust \citep[e.g.]{bang2005doubly} or double machine learning (DML) framework \citep{chernozhukov2016double}, we propose to correct the bias incurred by $\mlh-\bar{m}_l$ through the augmented estimation
\begin{align}
    \qlh & = \widetilde{\beta}_l + \cplh \wlh(X^{(l)}_i) [Y^{(l)}_i-\mlh(X^{(l)}_i)] (\szh)^{-1} A\ul_i,\quad\forall~l \in [L], \label{eq:qlhat} \\
    \widehat{\beta}_{\mix} & = \widetilde{\beta}_{\mix} + \suml \cplh \elh(X\ul_i) \wlh(X^{(l)}_i) [Y^{(l)}_i-\mlh(X^{(l)}_i)] (\szh)^{-1} A\ul_i := \ph, \label{eq:ql+1hat}
\end{align}
where $\wlh(x)$ is an estimator of the density ratio between the target and source $l$: $\bar{w}_l(x) = dP\xz(x)/dP\xl(x)$ obtained using general learning methods. For high-dimensional $A$, we can naturally extend this doubly robust construction to a regularized regression version with penalties like ridge or Lasso. See the detailed forms in Appendix.

Essentially, $\beta_l$'s are parameters defined on some hypothetical (counterfactual) populations without actual samples due to covariate shift. Thus, our case falls into the semiparametric estimation problem frequently studied in the contexts of missing data and causal inference. Inspired by existing semiparametric literature \citep[e.g.]{robins1994estimation}, we establish in Proposition \ref{propo:Doubly} the doubly robust property of $\qlh$ with respect to the two sets of nuisance models $w_l$ and $m_l$ for $l\in[L+1]$.
\begin{proposition}
\label{propo:Doubly}
Denote by $\qlh = \qlh(\wlh,\mlh), l \in [L]$ and $\widehat{\beta}_{\mix} = \widehat{\beta}_{\mix}(\elh,\wlh,\mlh)$ in (\ref{eq:qlhat}) and (\ref{eq:ql+1hat}). We have that $\qlh(w_l,m_l)$ is consistent for $ \bar{\beta}_l$ and $ \widehat{\beta}_{\mix}(\bar{\eta}_l,w_l,m_l)$ is consistent for $\bar{\beta}_{L+1}$ when either $w_l = \bar{w}_l$ or $m_l = \bar{m}_l$ holds for every $l \in [L]$.
\end{proposition}

Proposition \ref{propo:Doubly} implies that our constructions in (\ref{eq:qlhat}) and (\ref{eq:ql+1hat}) are less sensitive and doubly robust to the misspecification or excessive errors in the source-target density ratio model $w_l$ and the conditional mean model $m_l$. More detailed and rigorous quantification on this fact will be given in Theorem \ref{thm:betarate}. In our primary setup with the semi-supervised structure $N_l\gg n_l$ and larger source samples $N_l\gg N_0$ for $l\in[L]$, our estimation procedure of $\eta_l$ to be introduced in Section \ref{sec:method:est:nuis} can include larger samples for machine learning and, thus, produce an $\elh$ converging substantially faster than $\wlh$ and $\mlh$ when these models are of similar complexity. One can find more details in Section \ref{sec:theory}.

After obtaining $\widehat{\beta}_l$ for $l\in[L+1]$, we can directly use Theorem \ref{thm:PopuIdenti} to derive the DORM estimator with some pre-specified $\smax$. In specific, we compute $\widehat\Gamma=(\widehat\Gamma_{l,k})_{l,k \in [L+1]}$ with $\widehat\Gamma_{l,k}=\widehat{\beta}_l\trans \widehat\Sigma\uz \widehat{\beta}_k$ and $\widehat\Sigma\uz=\cpzh A^{(0)}(A^{(0)})\trans$, then solve the quadratic programming problem for the adversarial weights $\widehat{\gamma}= \arg\min_{\gamma \in \mathcal{S}(\smax)} \gamma\trans\widehat{\Gamma}\gamma$ and obtain $\ours(\smax) = \sum_{l=1}^{L+1}\glh \qlh $. In our setup, estimating the nuisance functions in (\ref{eq:btil}), (\ref{eq:qlhat}) and (\ref{eq:ql+1hat}) is not a straightforward application of machine learning considering the source-mixture structure and the relatively small size of the target data compared to the sources. Next, we shall propose a novel pipeline for efficient estimation of $\bar{\eta}_l$, $\bar{w}_l$ and $\bar{m}_l$ allowing the use of general machine learning methods.

\subsection{Estimation of nuisance models}\label{sec:method:est:nuis}

The conditional mean model estimator $\mlh$ can be obtained in a straightforward way, regressing $Y\ul\sim X\ul$ on the $n_l$ labeled sample in source $l$, using arbitrary learning methods such as Lasso, random forest, and neural networks. To estimate $\bar{\eta}_l$ in (\ref{eq:y|xmix}) with unknown $\bar\rho_l$ defined by (\ref{eq:kldiv}), one may consider first estimating each $d\PP^{(l)}_X$ for $l\in[L]$, then plugging them into the empirical version of (\ref{eq:kldiv}) on the target sample to estimate $\bar\rho_l$, and using the estimators of $d\PP^{(l)}_X$ and $\bar\rho_l$ for construction. However, this procedure may result in poor accuracy due to the common difficulty in estimating potentially high-dimensional density functions like $\PP^{(l)}_X$ \citep{gretton2009covariate}. 

To avoid this issue, our idea is to leverage the fact that
\begin{equation}
\bar\eta_l(x)=\frac{\bar\rho_l \bar{r}_l(x)}{\sum_{k=1}^L\bar\rho_k \bar{r}_k(x)};\quad   \bar\rho= \max_{\rho \in \Delta_L}\EE_{\PP_X^{(0)}}\log\Big(\sum_{l=1}^L \rho_l \bar{r}_l(x)\Big),    
\label{eq:tildedensityratio}
\end{equation}
where $\bar{r}_l(x)$ is some density ratio function
\[
    \bar{r}_l(x):=\frac{d \mathbb{P}_X^{(l)}(x)}{d\PP_X^{(\operatorname{ref})}(x)} =  \frac{d \mathbb{P}_X^{(0)}(x)}{\bar{w}_l(x)d \PP_X^{(\operatorname{ref})}(x)},\quad l\in[L],
\]
and $\PP_X^{(\operatorname{ref})}$ is an arbitrary reference distribution with its domain covering each source $l$ and will be specified later. Based on (\ref{eq:tildedensityratio}), we can replace the need in estimating $d\PP^{(l)}_X$ with $\bar{r}_l(x)$, which can be converted to a classification problem. In specific, for each $l \in[L]$, we merge $\{X_i^{(l)}\}_{i \in [N_l]}$ with samples from the reference distribution $\{X_i^{(\operatorname{ref})}\}$ and denote an observation randomly drawn from this pooled data set as $\tilde{X}_i^{(l)}$. Then define a response $G_i^{(l)} \in\{0,1\}$ such that $G_i^{(l)}=1$ if $\tilde{X}_i^{(l)}$ is actually from source $l$ and $G_i^{(l)}=0$ if $\widetilde{X}_i^{(l)}$ is from the reference data set. Then we have $d \mathbb{P}_X^{(l)}(x)=d\PP(x \mid G_i^{(l)}=1)$ and $d \mathbb{P}^{(\operatorname{ref})}_X(x)=d\PP (x \mid G_i^{(l)}=0)$ and that
\begin{equation}
    \bar{r}_l(x)=\frac{\PP(G_i^{(l)}=0)}{\PP (G_i^{(l)}=1)} \frac{\PP(G_i^{(l)}=1 \mid \tilde{X}_i^{(l)}=x)}{\PP(G_i^{(l)}=0 \mid \tilde{X}_i^{(l)}=x)}.
\label{eq:densityratio}
\end{equation}
In (\ref{eq:densityratio}), $\PP (G_i^{(l)}=0)/\PP (G_i^{(l)}=1) $ can be approximated by the sample size ratio between the reference data and source $l$, and $\PP (G_i^{(l)}=1 \mid \tilde{X}_i^{(l)}=x)$ can be estimated using general machine learning methods such as logistic Lasso and random forest classifier on the full source samples with sizes $N_l$'s. When $\PP_X^{(\operatorname{ref})}$ is properly specified, such a classification task is typically much easier to implement and has more accuracy compared to the direct estimation of $d \mathbb{P}_X^{(l)}(x)$. Alternatively, one could also estimate $\bar{r}_l$ using covariate balancing approaches such as \cite{gretton2009covariate} and \cite{imai2014covariate}. 
\begin{remark}
For the sake of estimation efficiency, the reference sample $\{X_i^{(\operatorname{ref})}\}$ can be taken as either (i) the source site with the largest sample size or (ii) the pool of all source samples, to maximize the size of training data. Choice (ii) also ensures sufficient overlap between $\PP_X^{(\operatorname{ref})}$ and every $\PP_X^{(l)}$ and, consequently, more stable training with general machine learning. Empirical estimation under choice (ii) can be realized by randomly splitting each $\Dsc_l$ into two sets with certain proportions, pooling one of them together for $l\in[L]$ to form the reference data, and learning $\bar{r}_l(x)$ by classifying the other set versus the pooled reference data. 
\end{remark}
With the estimator $\widehat{r}_l(x)$, we can then estimate $\rho$ through
\begin{equation}
    \widehat{\rho} = \underset{\rho \in \Delta_L}{\arg\max}\ \cpzh \log\Big(\sum_{l=1}^L\rho_l \widehat{r}_l(x)\Big)+\lambda\|\rho\|_2^2,
\label{eq:kldiv-est}
\end{equation}
and obtain the posterior weights as 
\begin{equation}
    \elh(x) = \frac{\widehat{\rho}_l{\widehat{r_l}(x)}}{\sum_{k=1}^L \widehat{\rho}_k\widehat{r}_k(x)}.
\label{eq:etalhat}
\end{equation}
In (\ref{eq:kldiv-est}), the ridge penalty on $\rho$ is introduced with a small $\lambda$ to secure a unique and stable solution when some sources $l$ and $k$ have extremely close density functions of $X$ and highly correlated $\widehat{r}_{l}(X)$ and $\widehat{r}_{k}(X)$ on the target sample. For the density ratio between source $l$ and target $\bar{w}_l$, we provide two options on its estimation. The first one is to use a similar procedure as introduced above to obtain a general machine learning estimate $\widehat{r}_0(x)$ of ${d \mathbb{P}_X^{(0)}(x)}/{d \PP_X^{(\operatorname{ref})}(x)}$ with the target and reference sample, and take each $\widehat{w}_l(x)=\widehat{r}_0(x)/\widehat{r_l}(x)$. When the target has a relatively small sample size $N_0$ compared to the sources, the error of $\widehat{w}_l$ tends to be dominated by $\widehat{r}_0(x)$ and the small $N_0$.

This motivates us to propose an alternative option by trusting the mixture form of $\PP_X^{(0)}$ in (\ref{eq:xmix}) and taking advantage of the mediating results in estimating $\widehat{\eta}_l$. In specific, one could use the mixture of density ratio $\sum_{l=1}^L\widehat\rho_l \widehat{r}_l(x)$ fitted from (\ref{eq:kldiv-est}) to approximate $\widehat{r}_0(x)$ and again take $\widehat{w}_l(x)=\widehat{r}_0(x)/\widehat{r_l}(x)$. This strategy can alleviate the small $\Dsc_0$ issue because given $\widehat{r_l}(x)$ estimated using the relatively large source data sets, (\ref{eq:kldiv-est}) is an $M$-estimation problem only involving $L$ parameters in $\rho$ rather than the complex models for $X$, which typically attains a parametric convergence rate with respect to the sample size $N_0$; see Lemma \ref{lemma:eta} for more details. However, this approach is subject to potential bias caused by the misspecification of (\ref{eq:xmix}), which is not an issue of our first option.

Inspired by \cite{chernozhukov2016double}, we also use cross-fitting in our framework to remove excessive bias caused by the complicated machine learning estimation of the nuisance models. For simplicity, notations related to this procedure are omitted in the main paper. We shall shortly introduce it here with the complete details presented in Appendix. For all the sources $l\in[L]$, we randomly partition $\Dsc_l$ into two disjoint subsets $\mathcal{A}_l$ and $\mathcal{B}_l$ with equal size, and construct two sets of estimators $\{\wlha,~\elha,~\mlha:l\in[L]\}$ and $\{\wlhb,~\elhb,~\mlhb:l\in[L]\}$ separately using $\Asc=\cup_{l=1}^L\mathcal{A}_l$ and $\Bsc=\cup_{l=1}^L\mathcal{B}_l$, with the whole target sample $\Dsc_0$ used for both sets. Denote the sub-sample empirical operators on each $l\in[L]$ as $\cplha = |\Asc_l|\inv \sum_{i \in \Asc_l}$ and $\cplhb = |\Bsc_l|\inv \sum_{i \in \Bsc_l}$. Then we construct $\qlha$ for $l\in[L+1]$ by plugging $\cplha$ and $\{\wlhb,~\elhb,~\mlhb\}$ into (\ref{eq:btil}), (\ref{eq:qlhat}), and (\ref{eq:ql+1hat}), and $\qlhb$ plugging in $\cplhb$ and $\{\wlha,~\elha,~\mlha\}$. At last, we separately combine the two sets of estimators $\qlha$'s and $\qlhb$'s with the full sample $\widehat\Sigma\uz$ to derive the corresponding adversarial weights $\glha$'s and $\glhb$'s following Theorem \ref{thm:PopuIdenti}, and output $\ours(\smax) = \sum_{l=1}^{L+1} (\glha \qlha + \glhb \qlhb)/2$ as the final DORM estimator.

\subsection{Tuning with side information}\label{sec:tuning}

The only hyper-parameter $\smax$ plays an important role in our method as it controls the size of the uncertainty set $\Msc(\smax)$ and the degree of adversary. Usually, choosing a relatively small $\smax$ (says $0.1$) to ensure enough weights on the source-mixing estimator $\widehat{\beta}_{\mix}$ is highly recommended when the key assumption (\ref{eq:ideal joint}) is reasonable, e.g., our real-world application targeting mixed ethnicity subgroups. Moreover, when there is some side information about outcome $Y^{(0)}$ on the target, we can potentially use it to find a good choice on $\smax$. We shall make some discussions and proposals about this point here and further justify them in both theoretical and numerical studies.

Suppose there is a small set of labeled data $(X_i^\dagger,Y_i^\dagger)$ for $i \in [n_\dagger]$ where $n_\dagger$ could be much smaller than both $N_0$ and $n_l$'s. The distribution of these labeled data used for tuning could come from another similar study or from multiple institutions, which is allowed to be moderately different from the specific target $(X\uz,Y\uz) \sim \PP\uz$. In specific, $(X_i^\dagger,Y_i^\dagger) \sim \PP^\dagger$ can be some distribution in the previous uncertainty set $\uncer$ in (\ref{eq:uncertainty}), or with some contamination part as in (\ref{eq:contam1}). See Theorem \ref{thm:truerate} in Section \ref{sec:theory} and the real data analysis in Section \ref{sec:realdata} for more explanation and examples. With such a limited sample size frequently encountered in practice, a potentially high-dimensional regression for $Y_i^\dagger\sim A_i^\dagger$ may not be tractable. However, we can use these labeled samples for the tuning of $\smax$, which tends to have a much lower complexity than the regression. With a series of fitted $\ours(\smax)$ for $\smax$ from some candidate set $\Csc$ (e.g., a uniform grid on $[0,0.5]$), we choose the one with the smallest mean prediction error on the labeled data, i.e., $\ours(\widehat{\smax})$ with
\begin{equation}
    \widehat{\smax} = \underset{\smax\in\Csc}{\arg\min}\   \frac{1}{n_\dagger} \sum_{i=1}^{n_\dagger} \big\{Y^\dagger_i-\ours(\smax)\trans A^\dagger_i\big\}^2.
    \label{eq:tuning}
\end{equation}

\begin{remark}
$1-\widehat{\smax}$ obtained by (\ref{eq:tuning}) can be viewed as an estimate of the optimal weight assigned to the source-mixing model coefficient $\bar{\beta}_{\mix}$, with the purpose of optimizing the prediction performance on the target domain among all DORM estimators with $\smax\in\Csc$. As will be justified in Theorem \ref{thm:truerate}, when the source-mixing assumption (\ref{eq:ideal joint}) approximately holds (with small contamination), this tuning procedure is ensured to result in no worse prediction performance compared to $\bar{\beta}_{\mix}$ derived as the optimal choice under (\ref{eq:ideal joint}).    
\end{remark}

When the outcome $Y$ is strictly unavailable for the target, one could instead use a surrogate outcome \(S_i^\dagger\) as a noisy approximation to guide model selection or tuning. For instance, in an EHR-based application, \(S_i^\dagger\) might be the counts of relevant diagnostic codes that approximate the disease outcome; in a clinical study, \(S_i^\dagger\) might represent an early endpoint or biomarker that serves as a proxy for the long-term outcome of primary interest. In both cases, the surrogate outcome is typically easier and less expensive to collect than the actual outcome.

As shown in previous work like \cite{li2023maxway}, under the common used conditional independence assumptions such as 
\[
S_i  \,\perp\, A_i  \,\big\vert\, Y_i 
\quad\text{or}\quad
Y_i  \,\perp\, A_i  \,\big\vert\, S_i ,
\]
the linear model \(S_i  \sim A_i \) shares the same direction of the coefficients as \(Y_i  \sim A_i \). Intuitively, if \(S_i \) tracks \(Y_i \) reasonably well (even with noise), a coefficient vector \(\beta\) that predicts \(Y_i \) will also tend to induce non-trivial correlation between \(S_i \) and \(\beta^\top A_i \). Concretely, under these conditions, having \(\beta\) predictive of \(Y_i \) implies a large (in absolute value) empirical correlation \(\widehat{\text{Cor}}(S , \beta\trans A )\) on the target domain. 

Based on this reasoning, in the absence of $Y$'s observation on the target domain, we can choose the hyperparameter \(\smax\) that optimizes a correlation-based criterion involving the surrogate outcome. Specifically, define
\[
\widehat{\smax} \;=\; \underset{\smax \in \Csc}{\arg\min}
\;\widehat{\text{Cor}} \bigl(S^\dagger,\,\ours(\smax)\trans A^\dagger\bigr),
\]
where \(\widehat{\text{Cor}}\) denotes the empirical correlation measured on the target sample. By maximizing the correlation with the surrogate, we harness its partial alignment with the true outcome while acknowledging that \(S_i^\dagger\) is only an approximation. This procedure thereby offers a practical and statistically motivated way to tune \(\smax\) even when no ground-truth outcomes are available for the target population.

\section{Theoretical Justification}\label{sec:theory}

In this section, we investigate the convergence property of $\ours(\smax)$ and the effectiveness of the tuning procedure in Section \ref{sec:tuning}. For two real numbers $a,b$, we denote by $a \wedge b := \min\{a,b\}$. For two sequences $a(n)$ and $b(n)$, we use $a(n) \lesssim b(n)$ or $a(n)=O(b(n))$ to represent that there exists some universal constant $C>0$ such that $a(n) \leq C b(n)$ for all $n \geqslant 1$, and use $a(n) \lesssim_{\sf P} b(n)$ or $a(n)=O_{\sf P}(b(n))$ for $a(n) \lesssim b(n)$ or $a(n)=O(b(n))$ with a probability approaching $1$. For some matrix $B$, we use $\lambda_{\min}(B)$ and $\lambda_{\max}(B)$ to denote its smallest and largest eigenvalues respectively. We use $N_{[L]}$ to denote all the unlabeled sample sizes $\{N_1,N_2,\cdots,N_L\}.$ On the target and sources we denote expectation operators by $\EE_0 = \EE_{\PP\uz_X}$ and $\EE_l=\EE_{(\PP_X\ul, \PP_{Y|X}\ul)}$. 

In addition, we denote the density ratio with respect to the reference group as $\bar{r}(X) := (\bar{r}_1(X),\cdots,\bar{r}_L(X))\trans$, $\widehat{r}(X) = (\widehat{r}_1(X),\cdots,\widehat{r}_L(X))\trans$, and the Hessian matrix of $\bar{\rho}$ in (\ref{eq:tildedensityratio})
\[
\bar\Omega(\rho)=\big(\bar\Omega_{lk}(\rho)\big)_{l,k\in[L]},\quad\mbox{with}\quad \bar\Omega_{lk}(\rho)=\EE_0\Bigg[\frac{\bar{r}_l(X)\bar{r}_k(X)}{\big(\sum_{i=1}^L\rho_i\bar{r}_i(X)\big)^2}\Bigg].
\]
Set the penalty coefficient $\lambda =O( N^{-1/2}_0)$ in (\ref{eq:etalhat}). For some measurable function $f: \mathbb{R}^p \to \mathbb{R}$, we define its $A$-norm and $\ell_2$-norm over the target and source covariate distribution as 
\begin{equation}
    \|f\|_{A,l} := \big(\EE_l \| f(X) A\|_2^2\big)^{1/2},\quad l = 0,1,\cdots,L; \label{eq:A-norm}
\end{equation}
\begin{equation}
    \|f\|_{2,l} := \big(\EE_l f(X)^2\big)^{1/2},\quad l = 0,1,\cdots,L.
    \label{eq:L2norm1}
\end{equation}
If $f$ is a vector function $f:\mathbb{R}^p \to \mathbb{R}^{p'}$, we extend the definition of $\ell_2$-norm to
\begin{equation}
    \|f\|_{2,l} := \big(\EE_l \|f(X)\|_2^2\big)^{1/2},\quad l = 0,1,\cdots,L.
    \label{eq:L2norm2}
\end{equation}
We now introduce and comment on our main assumptions as follows.

\begin{assumption}
There exists some constant $C_1>0$ such that $\EE_lY^2<C_1$, $C_1\inv \leqslant \lambda_{\min}(\bar\Sigma\uz) \leqslant \lambda_{\max}(\bar\Sigma\uz) \leqslant C_1$, and $C_1\inv \leqslant \lambda_{\min}(\bar\Omega(\bar\rho)) \leqslant \lambda_{\max}(\bar\Omega(\bar\rho)) \leqslant C_1$. Covariates $A\uz$ has finite fourth moments $\EE_0 \|A\|_2^4 < \infty$.
\label{assum:1}
\end{assumption}

\begin{assumption}
\label{assum:bound}
There exists some absolute constant $C_2>0$ such that 
\begin{equation}
    \max\limits_{l \in[L],~k\in \{0,l\}} \| \bar{m}_l \|_{A,k}+{\|\bar{w}_l \|_{A,k} } \leqslant C_2,\quad \max\limits_{l \in [L+1]} \|\bar{\beta}_l\|_2 \leqslant C_2,\mbox{ where }~\bar{\beta}_{L+1}=\bar{\beta}_{\mix}.
\end{equation}
\end{assumption}

\begin{assumption}(Overlap) There exists some constant $C_w$ such that for $\forall l \in [L]$ and $x$, $0<C_w\inv < \wlb(x) < C_w$, $C_w\inv<\bar{r}_l(x)<C_w$.
\label{assum:overlap}
\end{assumption}

\begin{assumption}
\label{assum:rate} 
There exists sequences $\Delta^{r}(n)$, $\Delta^{w}(n)$ and $\Delta^{m}(n)$ that converge to $0$ as $n\rightarrow\infty$, and satisfy
\begin{align*}
\max\limits_{k\in \{0,l\}} \left\|\mlh - \bar{m}_l\right\|_{A,k} &\lesssim_{\sf P} \Delta^{m}(n_l);\\
\max\limits_{k\in \{0,l\}} \left\|\wlh - \bar{w}_l\right\|_{A,k} + \left\|\wlh - \bar{w}_l\right\|_{2,k} &\lesssim_{\sf P} \Delta^{w}(N_0\wedge N_l);\\
\max\limits_{k\in \{0,l\}} \left\|\widehat{r} - \bar{r}\right\|_{2,k}\ &\lesssim_{\sf P} \Delta^{r}(N_{[L]}).
\end{align*}
\end{assumption}

Assumptions \ref{assum:1} and \ref{assum:bound} are mild and common regularity and boundedness assumptions on the data distributions and models. The condition on $\bar\Omega(\bar\rho)$ is imposed to ensure proper convergence of $\widehat{\rho}_l$ in  (\ref{eq:etalhat}). In Assumption \ref{assum:overlap}, we assumed the upper bound and lower bound for density ratios $\bar{w}_l(x)$ and $\bar{r}_l(x)$, which could be understood as requiring $\PP_X$'s overlap between the sources and target or the reference group are not too weak. Similar positivity assumptions have been commonly employed in existing literature \citep{sugiyama2012density,liu2023augmented}. The constant $C_w$ characterizes the extent of this overlap, and we retain this constant in the subsequent convergence rate analysis to demonstrate how the rates depend on the degree of overlap between the sources and target.

Assumption \ref{assum:rate} imposes requirements on the error rates of the general machine learning estimators for the nuisance models, in a similar spirit as the DML framework \citep{chernozhukov2016double}. For simplification, we consider a slightly less general regime that the error rate functions of $n$ hold to be the same across all the source and target domains and the training sample size is sufficient for characterizing the convergence rates of the nuisance machine learners. For instance, we use $\Delta^{m}(n_l)$ to depict the error rate of $\widehat{m}_l$ as $\widehat{m}_l$ is estimated using the labeled sample from source $l$ with sample size $n_l$. An underlying assumption of this regime is that the nuisance models have the same complexity or structural property (e.g., sparsity and smoothness) across all sites.


\begin{remark}
Each $\Delta$ function in Assumption \ref{assum:rate} is supposed to converge faster to $0$ with a increased sample size under proper learning models. Such convergence properties have been well established for various methods such as Lasso \citep[e.g]{negahban2009unified}, random forest \citep[e.g.]{athey2019generalized}, and deep neural network \citep[e.g.]{farrell2021deep}. In all these examples, $\Delta$ has been shown to converge with a polynomial rate of the training sample size. Considering that the effective sample size of a classification problem is typically determined by its smaller class, we write $\Delta^{w}(N_0\wedge N_l)$ in Assumption \ref{assum:rate}. For the auxiliary density ratio $\widehat{r}$, since it is only used in estimating the prior mixture weights $\widehat{\rho}$, we express its overall convergence in terms of the $\ell_2$-norm of all $L$ density ratios. Its convergence rate is thus influenced by the total number of unlabeled samples, denoted by $\Delta^r(N_{[L]})$.
\end{remark}

In the following Lemma \ref{lemma:eta}, we analyze our construction procedures in (\ref{eq:kldiv-est}) and (\ref{eq:etalhat}) to derive $\elh$'s convergence rate based on the error rate of $\widehat{r}_l$. Based on this lemma, we further establish the convergence properties of our DML estimators $\widehat{\beta}_l$ in Lemma \ref{lemma:Rrate}, which is a key intermediate results for our main theorem about the convergence rate of $\ours(\smax)$.

\begin{lemma}
\label{lemma:eta}
Under Assumptions \ref{assum:1}--\ref{assum:rate}, we have
\begin{equation}
\max\limits_{k\in \{0,l\}} \left\|\widehat{\eta}_l - \bar{\eta}_l\right\|_{2,k} \lesssim_{\sf P} { \Delta^{\eta}(N_0,N_{[L]}) } := \frac{L}{\sqrt{N_0}} + L   \Delta^r(N_{[L]}),
\label{eq:rate:eta}
\end{equation}
where the term $\Delta^r(N_{[L]})$ is defined in Assumption \ref{assum:rate}.
\end{lemma}

The first term on the right hand side of (\ref{eq:rate:eta}) corresponds to the learning error of the optimal weights $\bar\rho$ on the target samples by (\ref{eq:kldiv-est}) and the second error term arises from the machine learning estimation of $\bar{r}_l$'s with the full source sample. In the setup of our primary interests, we have $N_0<N_l$ or $N_0\ll N_l$. Due to the high complexity of machine learning models with $X$, $\Delta^r(N_{[L]})$ is usually slower than the parametric rate of any $N_l$ while the target samples only contribute a parametric error rate in $N_0$ thanks to the low-dimensionality of $\rho$. As a result, there is not a term always dominating the other one in $\Delta^{\eta}(N_0,N_{[L]})$.

\begin{lemma}
\label{lemma:Rrate}
    Under Assumptions \ref{assum:1}--\ref{assum:rate}, we have
    \begin{align}
        \| \widehat{\beta}_l - \bar\beta_l \|_2 &\lesssim_{\sf P} \text{Err}_l:=\sqrt{\frac{1}{N_0\wedge n_l}} + \Delta^m (n_l)\Delta^{w}(N_0\wedge N_l),\quad l\in [L];\label{eq:errl}\\
        \|\widehat{\beta}_{\mix} - \bar\beta_{\mix} \|_2 &\lesssim_{\sf P} \text{Err}_{\mix}:=\sum_{l=1}^L  \bar\rho_l \Big(\sqrt{\frac{1}{N_0\wedge n_l}} + \Delta^{m} (n_l)\Delta^{w}(N_0\wedge N_l)\Big) +\Delta^{\eta}(N_0,N_{[L]}). \label{eq:errmix}
    \end{align}
    where $\Delta^m (n_l),\Delta^{w}(N_0\wedge N_l)$ come from the convergence assumption in Assumption \ref{assum:rate}, and $\Delta^{\eta}(N_0,N_{[L]})$ is as defined in Lemma \ref{lemma:eta}.
\end{lemma}

The term $\sqrt{{1}/(N_0\wedge n_l)}$ in each $\text{Err}_l$ can be viewed as the oracle error one would get when plugging the true (population) nuisance models in $\qlh(w_l,m_l)$ and $\widehat{\beta}_{\mix}(\eta_l,w_l,m_l)$. The machine learning errors in $w_l$ and $m_l$ show up in $\text{Err}_l$ in a form of their production, which has a similar spirit as the general DML theory \citep[e.g.]{chernozhukov2016double}. Though $\Delta^{\eta}(N_0,N_{[L]})$ still appears as a first-order error without the DML correction, it tends to converge faster than $\Delta^{m} (n_l)$ and $\Delta^{w}(N_0\wedge N_l)$ in our setup with $N_l$ larger than both $N_0$ and $n_l$. Thus, our DML construction can effectively improve the convergence performance by addressing the possibly slowest terms $\Delta^{m}(n_l)$ and $\Delta^{w}(N_0\wedge N_l)$.

\begin{theorem}
\label{thm:betarate}
Under Assumptions \ref{assum:1}--\ref{assum:rate}, we have that for any $\smax\in[0,1]$,
\begin{equation}
 \| \ours(\smax) - \oursbar(\smax) \|_2 \lesssim_{\sf P}  \max_{l \in [L+1]} \text{Err}_l   + \frac{L^{3/2}\max_{l \in [L+1]} \text{Err}_l}{\lambda_{\min} (\Gamma) }\wedge L^{1/2}d(\Ssc(\smax)),
 \label{equ:thm2}
\end{equation}
where error terms $\text{Err}_l$ are as given in Lemma \ref{lemma:Rrate} with $\text{Err}_{L+1} :=\text{Err}_{\mix}$, $\Gamma$ and $\Ssc(\smax)$ are defined in Theorem \ref{thm:PopuIdenti}, and the diameter of $\Ssc(\smax)$ is defined as $d(\Ssc(\smax))= \max_{\gamma,\gamma' \in \Ssc(\smax)} \| \gamma - \gamma'\|_2$.
\end{theorem}
Theorem \ref{thm:betarate} characterizes the empirical estimation error of $\ours(\smax)$. The term $\text{Err}_l$ derived in Lemma \ref{lemma:Rrate} encodes the impact of $\widehat{\beta}_l$'s error. The second term in (\ref{equ:thm2}) is due to the group adversarial learning. When there are some highly correlated $\bar\beta_l$ and $\bar\beta_k$, $\Gamma=(\bar{\beta}_l\trans \bar\Sigma\uz \bar{\beta}_k)_{l,k\in[L+1]}$ could be nearly singular and the consequently large $1/\lambda_{\min} (\Gamma)$ will in turn inflate the error of $\ours(\smax)$. Also note that $d(\Ssc(\smax))=0$ as $\smax=0$. When $\smax$ vanishes fast to $0$, the error due to the adversarial learning on $\Dsc_0$ may have lower impact on $\ours(\smax)$. 


Furthermore, we investigate the tuning procedure defined in \eqref{eq:tuning}, which relies on \(n_\dagger\) labeled samples to find \(\widehat{\smax}\). To formalize this, we introduce
\begin{equation}
    \PP^{\operatorname{con}}(s^*) :=\Bigg\{ \Big(\PP\uz_X, (1-s^*)\sum_{l=1}^L \eta_l(X;\bar\rho) \mathbb{P}_{Y \mid X}^{(l)} 
       \;+\; s^* \PP^{(\varepsilon)}_{Y|X}\Big):\  \|\bar{m}^{(\varepsilon)}\|_{A,0} < C_3 \Bigg\},
\label{eq:contam1}
\end{equation}
for some $s^*\in[0,1]$, as a class of contaminated mixture conditional distributions, where \(\bar m^{(\varepsilon)}\) is the conditional expectation of \(Y\) with respect to \(\PP^{(\varepsilon)}_{Y|X}\). We assume that the joint distribution on the target domain $\PP\uz$ and the distribution of labeled samples used for tuning $\PP^\dagger$ are both in this class. The first source-mixing part in the $Y|X$ conditional distribution corresponds to our ideal assumption \eqref{eq:ideal joint} but is down-weighted by the probability \(1 - s^{*}\), and \(\mathbb{P}^{(\varepsilon)}_{Y|X}\) is an arbitrary contamination distribution, with weight \(s^{*}\). Without further specifying \(\PP^{(\varepsilon)}_{Y|X}\), we only require a mild regularity condition that \(\|\bar{m}^{(\varepsilon)}\|_{A,0} < C_3\) for some constant \(C_3 > 0\). For this arbitrary contamination part, we emphasize that the labeled samples used for tuning may have a different distribution from the actual outcome on the target domain, that is, the contamination part \(\PP^{(\varepsilon)}_{Y|X}\) could be different between $\PP^\dagger$ and $\PP\uz$.

Below, we show in Theorem~\ref{thm:truerate} that the selected model \(\ours (\widehat{\smax})\) will approach the best linear approximation model for \(Y \sim A\) on the target domain.

\begin{theorem}\label{thm:truerate}
Suppose that Assumptions \ref{assum:1}--\ref{assum:rate} hold and both the target domain distribution and tuning samples distribution \(\mathbb{P}\uz,\PP^\dagger \in \PP^{\operatorname{con}}(s^*)\) with any \(s^* \in [0,1]\). Then we have
\begin{equation}
\| \ours (\widehat{\smax}) - \beta^* \|_2 
\;\lesssim_{\sf P}\; s^*+\frac{1}{\sqrt{n_\dagger}}+\text{Err}_{\mix},
\label{equ:thm3}
\end{equation}
where \(\displaystyle\beta^* 
= \underset{\beta \in \mathbb{R}^q}{\arg\min}\ 
   \EE_{\mathbb{P}\uz}\bigl(Y - A\trans \beta\bigr)^2\) 
is the best linear model for \(Y \sim A\) under \(\mathbb{P}\uz\), \(n_\dagger\) is the labeled target sample size, and \(\text{Err}_{\mix}\) is given in Lemma~\ref{lemma:Rrate}.
\end{theorem}

On the right-hand side of \eqref{equ:thm3}, the first error term \(s^*\) reflects the presence of an arbitrary contamination distribution weighted by \(s^*\). The second term, \(1/\sqrt{n_\dagger}\), represents the cost of selecting \(\smax\) based on the \(n_\dagger\) labeled samples and follows a parametric rate of convergence. The third term \(\text{Err}_{\mix}\) represents the estimation error when no contamination is present (i.e., \(s^*=0\)) and captures the remaining error in the uncontaminated case. This result indicates that as contamination vanishes, the dominant sources of estimation error are controlled by the inherent complexity of the model and the available sample sizes, regardless of whether the labeled data used for tuning come from exactly the same or partially different distributions.

\begin{remark}
    Besides the distribution class in (\ref{eq:contam1}), our theoretical results can be potentially extended to the jointly contaminated mixture distribution on the target domain:
\begin{equation}
\PP^{\operatorname{con}}(s^*) :=\Big\{ (1-s^{*}) \sum_{l=1}^L\rho_l^* (\mathbb{P}_{X}^{(l)},\mathbb{P}_{Y|X}^{(l)})+ s^{*}\mathbb{P}^{(\epsilon)}\Big\},\quad~s^*\in[0,1],\ \rho^*\in\Delta^L.
    \label{eq:contam2}
\end{equation}
Here, instead of introducing contamination in the conditional distribution $\PP_{Y|X}\uz$, we apply a contamination model to the joint distribution in (\ref{eq:contam2}). Under this formulation and mild regularity conditions for $\PP^{(\epsilon)}$, we can also derive similar results as in Theorem \ref{thm:truerate}.
\end{remark}

\section{Simulation Study}

\subsection{Data generation and benchmarks}\label{simu:benchmarks}

We conduct comprehensive simulation studies to evaluate our proposed DORM under various settings and compare it with existing methods. In Section \ref{subsec:simu:Landmix}, we focus on low-dimensional $A$ settings and will compare DORM with benchmarks under various circumstances. In Section \ref{subsec:simu:diffsmax} we will show the effectiveness of our tuning strategy for $\smax$ and insensitivity of DORM to moderate deviation of $\smax$ from the optimal choice. In Section \ref{subsec:simu:highd} we will highlight the improved performance of DORM through comparison with several recently developed multi-source domain adaptation methods with high-dimensional $A$.

We set $L=5,N_l=2000,n_l=500,N_0=2000$ by default if not specified purposely. The dimension of $X$ is $p = 200$, and for low dimensional settings we set $q = \dim A = 5$. The simulation data are generated as follows. First, at each source site $l \in [L]$, we generate $A\ul_i (i \in [N_l])$ i.i.d. from a Gaussian distribution $N(\mu_l,\sigma^2_A I_{q})$ and set the first element 1. For each $W$, the first five elements are calculated by $W_1 = k(A_1-A_3),W_2=k(A_2-A_4),W_3=kA_3,W_4=W_5=kA_4$, and we keep $k=0.3$ creating a moderate correlation between $A$ and $W$. The rest elements of $W$ are 0, and we add a small noise $N(0,0.1^2)$ to each element of $W$. For each source site we generate the outcome $Y$ using a linear model
\[
    Y\ul_i = \alpha_l\trans A\ul_i + \gamma_l\trans W\ul_i + \varepsilon\ul_i,
\]
where $i \in [n_l]$ and $\varepsilon\ul_i$ is i.i.d Gaussian $N(0,0.5^2)$. It is note worthy that on each source site, model $\PP\xl$ depends on $\mu_l$; model $\PP_{Y|X}\ul$ is largely determined by $\alpha_l$ and $\gamma_l$. To model both covariate shift and posterior drift, we design that $\mu_l,\alpha_l$ and $\gamma_l$ are different across all the source sites. The detailed choice of $\mu_l,\alpha_l$ and $\gamma_l$ can be found in Appendix. 

For each subject $i$ on the target domain ($i \in [N_0]$), we generate $X\uz_i (i \in [N_0]) \sim \PP\uz_X = \suml \rho^*_l \PP\ul_X$ according to a given prior probability $\rho^* \in \Delta_L$, which will be detailed in the following experiments. For $Y\mid X$ on the target domain, we set the unobserved $Y\uz_i$ to follow the conditional distribution defined in (\ref{eq:Y|Xdist}):
\[ Y\uz_i | X\uz_i \sim \PP\uz_{Y|X} = (1-s^*)\sum_{l=1}^L \eta_l(X;\rho^*) \mathbb{P}_{Y \mid X}^{(l)}+ s^* \sum_{l=1}^L \delta_l^* \mathbb{P}_{Y \mid X}^{(l)}, \]
where $\delta^* \in \Delta^L$ is an arbitrary probability vector. Here the parameter $s^*$ characterizes the violation of the generated data from the ideal joint source-mixing assumption (\ref{eq:ideal joint}). The closer \(s^*\) is to 1, the farther the actual target distribution deviates from this source-mixing assumption.

We evaluate an estimator $\beta$ through its average or worst-case prediction performance on the uncertainty set of target domain using standardized MSE as the measure. To evaluate distribution robustness and generalizability, we uniformly sample 100 different $\delta^{*(b)}$ on the simplex $\Delta_L$ to generate the corresponding outcome $Y_i^{(0,b)}$, and report the average or worst-case performance (highest standardized MSE) among the 100 trials ($b=1,2,\cdots,100$). MSE is standardized by the average variance of $Y_i^{(0,b)}$.

By default, we generate a small $n_\dagger = 20$ labeled sample with $Y^\dagger_i$ to obtain $\widehat{\smax}$ using tuning method in Section \ref{sec:tuning}. We repeat 500 times of simulation for each setting to summarize the numerical performance. Our benchmarks include different aggregation strategies of the estimator $\qlh$ in (\ref{eq:qlhat}) and several recently developed high-dimensional domain adaptation approaches that will be mainly compared in Section \ref{subsec:simu:highd}. We summarize the methods under comparison as follows.
\begin{enumerate}
    \item \textbf{SimpleAve}: the simple average of each $\qlh$, $\widehat{\beta}_{\sf SA} = 1/L \suml \qlh$.
    \item \textbf{RhoAve}: the average of $\qlh$ weighted by $\widehat{\rho}$ obtained in (\ref{eq:kldiv-est}), $\widehat{\beta}_{\sf RA} = \suml \widehat{\rho_l} \qlh$.
    \item \textbf{Maximin}: the pure maximin regression by setting $\smax=1$ in our method, which can be viewed as a simple adaptation of the covariate shift adjusted maximin regression proposed in \cite{guo2022statistical} to our setup.
    \item \textbf{TransLasso}: Transfer Lasso approach proposed by \cite{li2022transfer}.
    \item \textbf{TransGLM}: Transfer Lasso for generalized linear models proposed by \cite{tian2023transfer}, with a different multi-source aggregation strategy from TransLasso.
    \item \textbf{PTL}: Profile Transfer Learning proposed by \cite{lin2024profiled}.
\end{enumerate}

\subsection{Performance under different $L$ and mixture weights}
\label{subsec:simu:Landmix}
In this subsection, we focus on the low-dimensional $A$ setting ($\dim A =5$) and change the prior mixture weights $\rho^*$ and the number of source sites $L$ to give a comprehensive result of our DORM method under different circumstances.

\begin{figure}[htbp]
\centering
\begin{minipage}[c]{0.48\textwidth}
  \includegraphics[width=\textwidth]{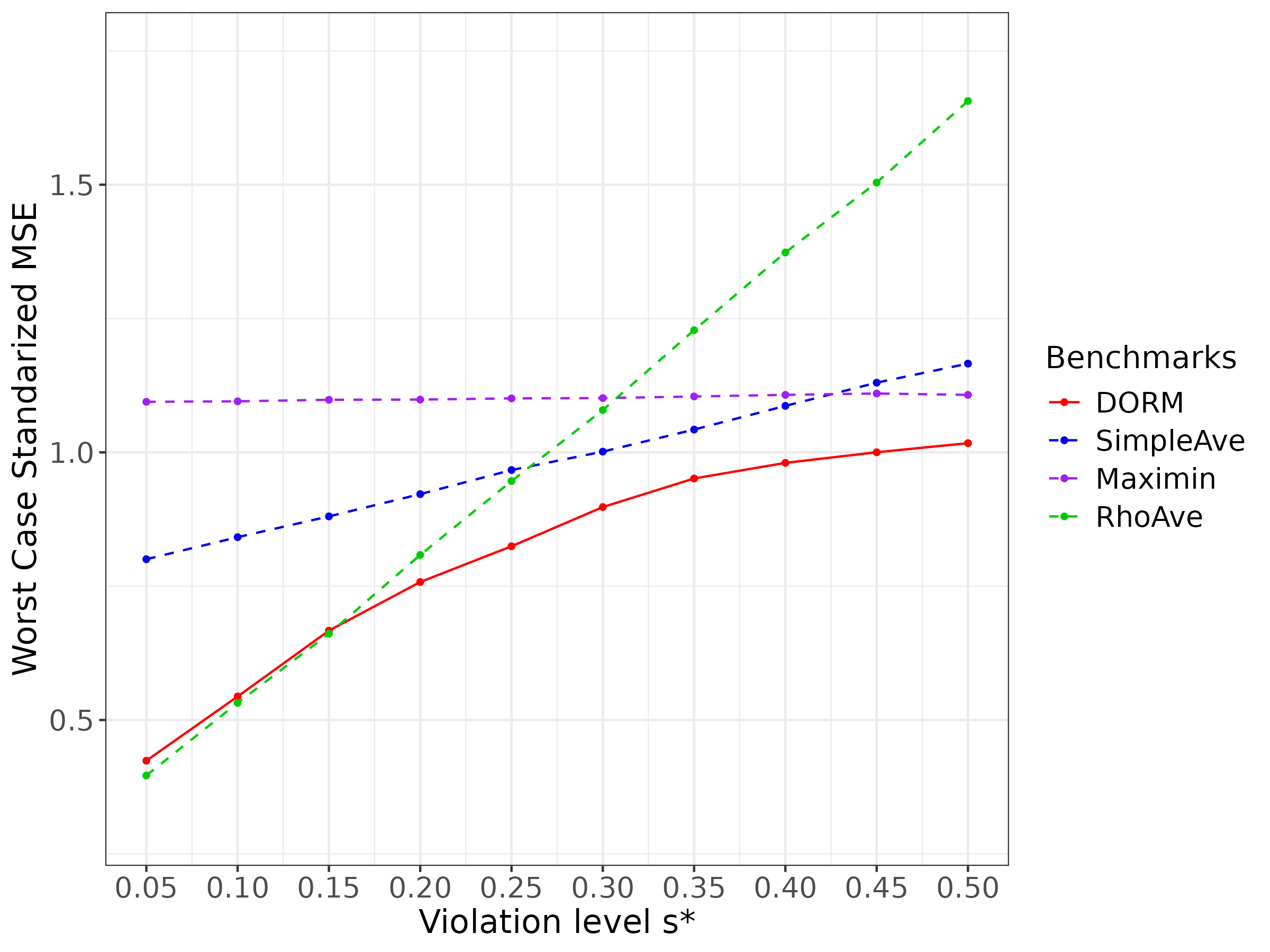}
  \subcaption{prior $\rho^*_1 = 0.5, \rho^*_3 = 0.5$}
  \label{fig:mix0}
\end{minipage}
\begin{minipage}[c]{0.48\textwidth}
  \includegraphics[width=\textwidth]{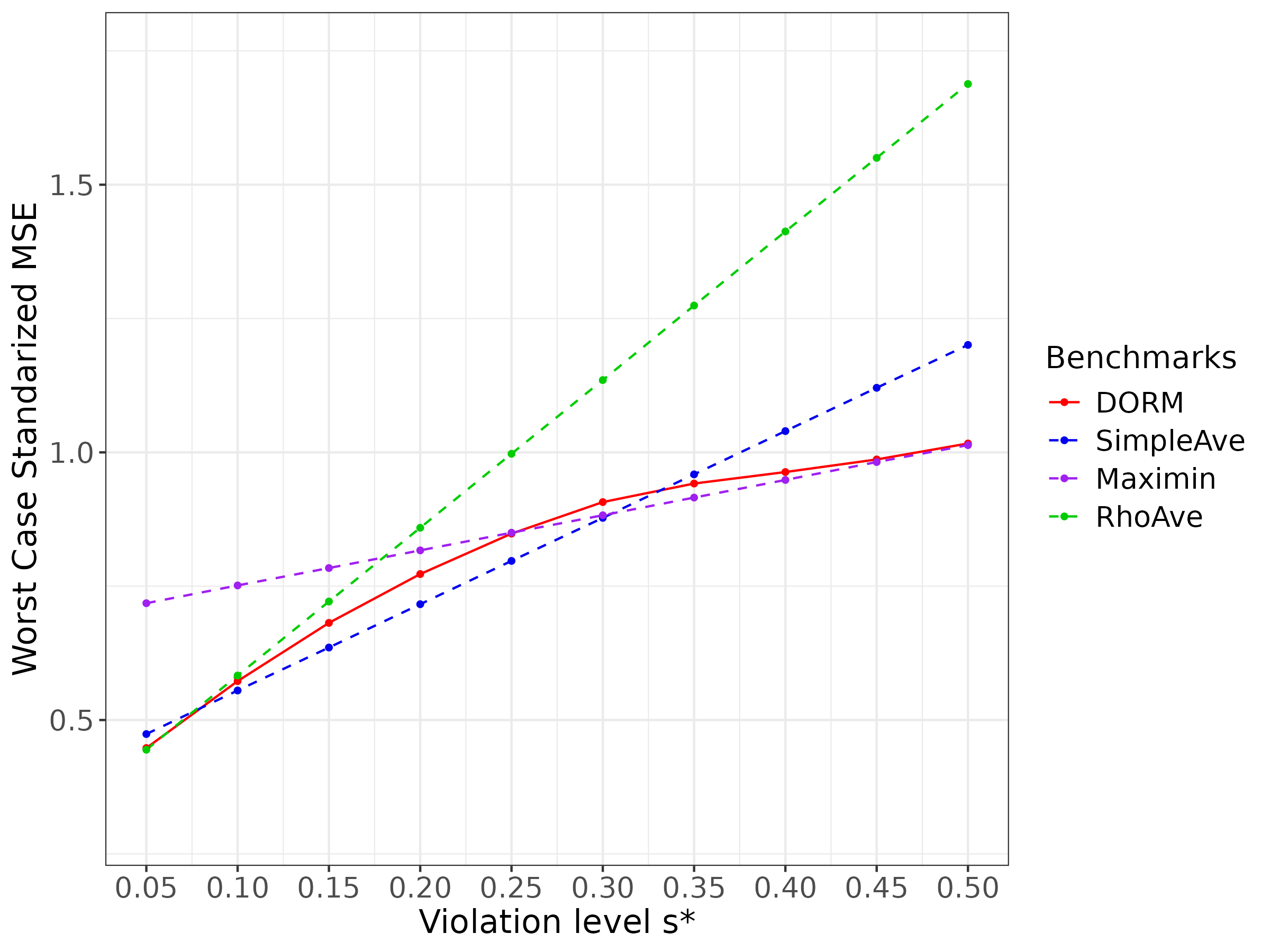}
  \subcaption{prior $\rho^*_1 = 0.8, \rho^*_3 = 0.2$}
  \label{fig:mix1}
\end{minipage}
\begin{minipage}[c]{0.48\textwidth}
  \includegraphics[width=\textwidth]{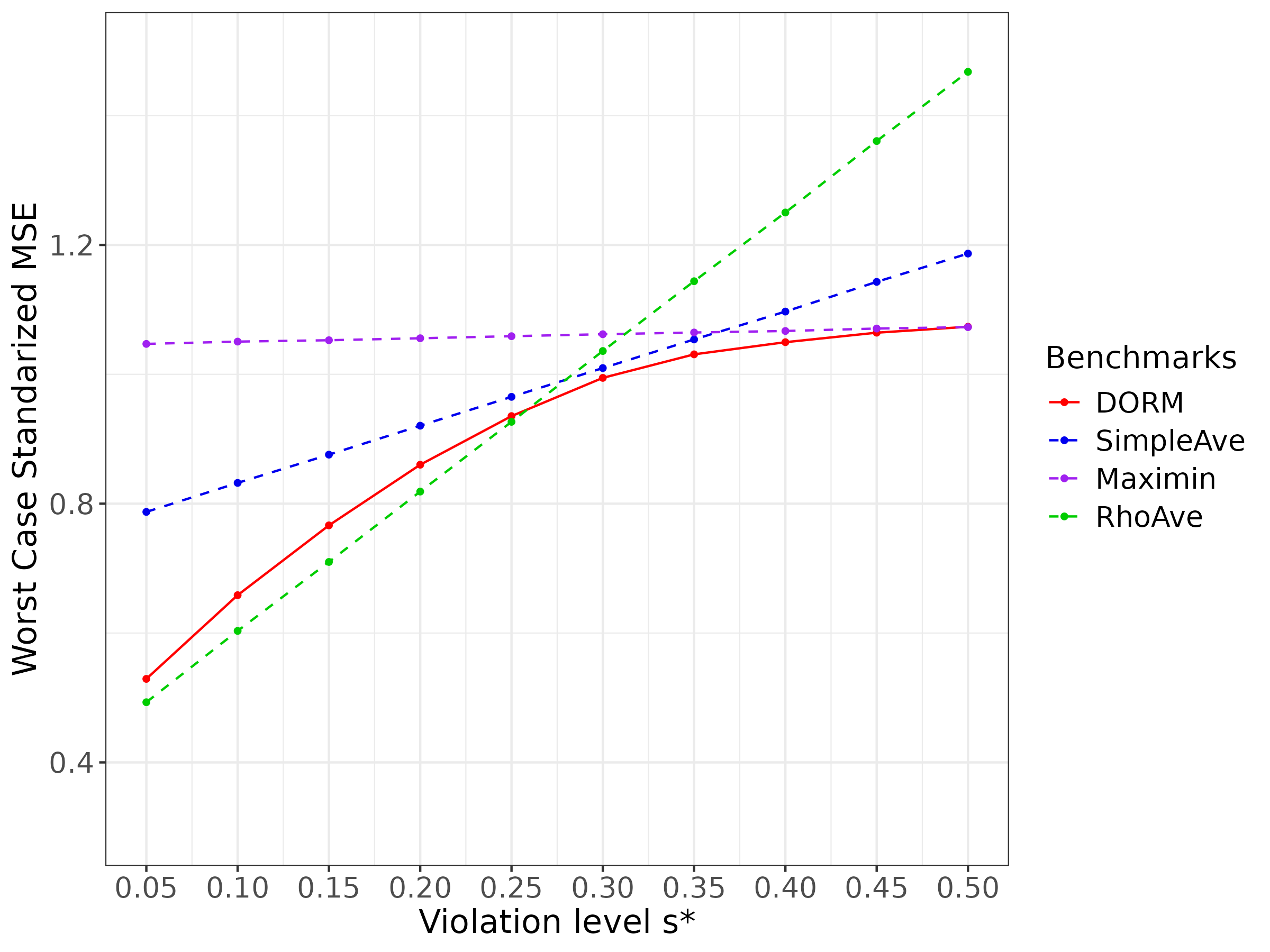}
  \subcaption{prior $\rho^*_1 = \rho^*_2 = \rho^*_3 = 1/3$}
  \label{fig:mix2}
\end{minipage}
\begin{minipage}[c]{0.48\textwidth}
  \includegraphics[width=\textwidth]{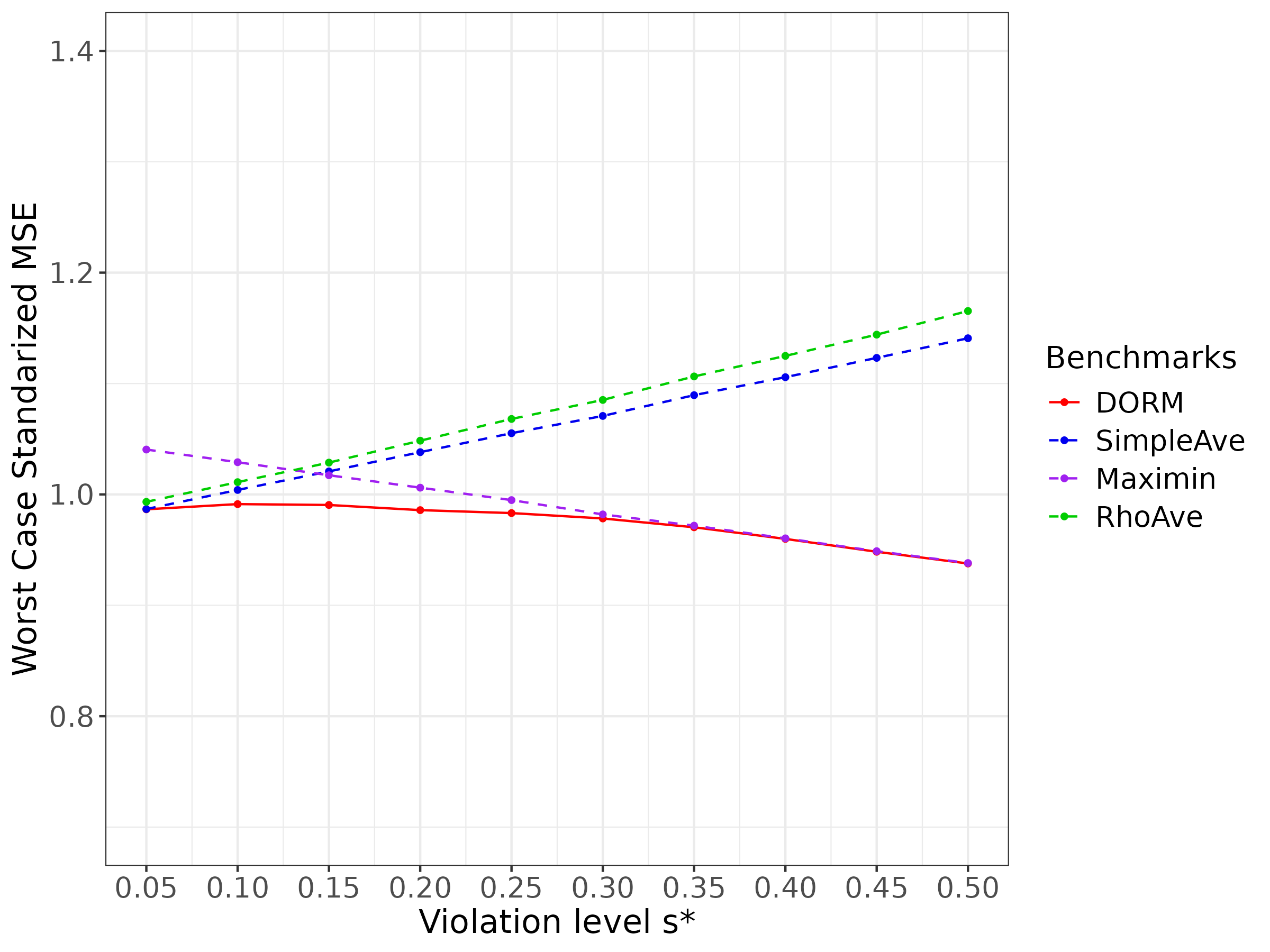}
  \subcaption{prior: equal weight 1/5 }
  \label{fig:mix3}
\end{minipage}
\caption{The worst case performance with different mixture structures. In the first panel the prior mixture of $X$ is set to be $\rho^* = (0.5,0,0.5,0,0)$. In the second panel, the prior mixture is $\rho^* =(0.8,0,0.2,0,0)$; in the third panel, the prior mixture is $\rho^* =(1/3,1/3,1/3,0,0)$; in the last panel, the prior mixture is $\rho^* =(0.2,0.2,0.2,0.2,0.2)$. In all plots, the red solid line is our method DORM with tuned $\widehat{\smax}$, and four dashed lines are benchmarks.}
\label{fig:diff-mix}
\end{figure}

In Figure \ref{fig:diff-mix} we have $L=5$ source sites and change the prior probability $\rho$ to different settings. We report the worst-case performance under different violation levels $s^* \in [0,0.5]$. In the first scenario \ref{fig:mix0}, the prior mixture is $\rho^* = (0.5, 0, 0.5, 0, 0)$, which means we will choose the target from source site 1 or 3 with equal probability. We can observe that under most of violation levels, DORM consistently outperforms all the benchmarks. When $s^*$ is very small ($s^* \in [0.05,0.20]$), our method is very close to RhoAve, which is reasonable because our method is designed to handle robustness against the violation of ideal mixture. In the meantime DORM performs much better than SimpleAve and Maximin. When there is moderate violation, for example $s^* = 0.35$, standardized MSE of DORM is 0.9412, which is 10.77\% better than SimpleAve, 17.34\% better than Maximin and 30.49\% better than RhoAve. RhoAve gives a poor performance when $s^*$ is large because the mixing weights $\rho^*$ determined by $X$ have been violated. 

In the second scenario \ref{fig:mix1}, the prior is more unbalanced, with more weight on the first site and less weight on the third site. We can see that our method still outperforms most benchmarks at a moderate $s^*$. We notice that Maximin is almost as good as our method in this unbalanced scenario, which means a moderate uncertainty set has already arrived at the same optimal coefficient as the maximum uncertainty set. In the third scenario \ref{fig:mix2}, we have a balanced mixture among the first three sites. The phenomenon is almost the same as Figure \ref{fig:mix0}, showing that equal mixture of two sites and three sites are similar. 

\begin{figure}[htbp]
\centering

  \includegraphics[width=0.5\textwidth]{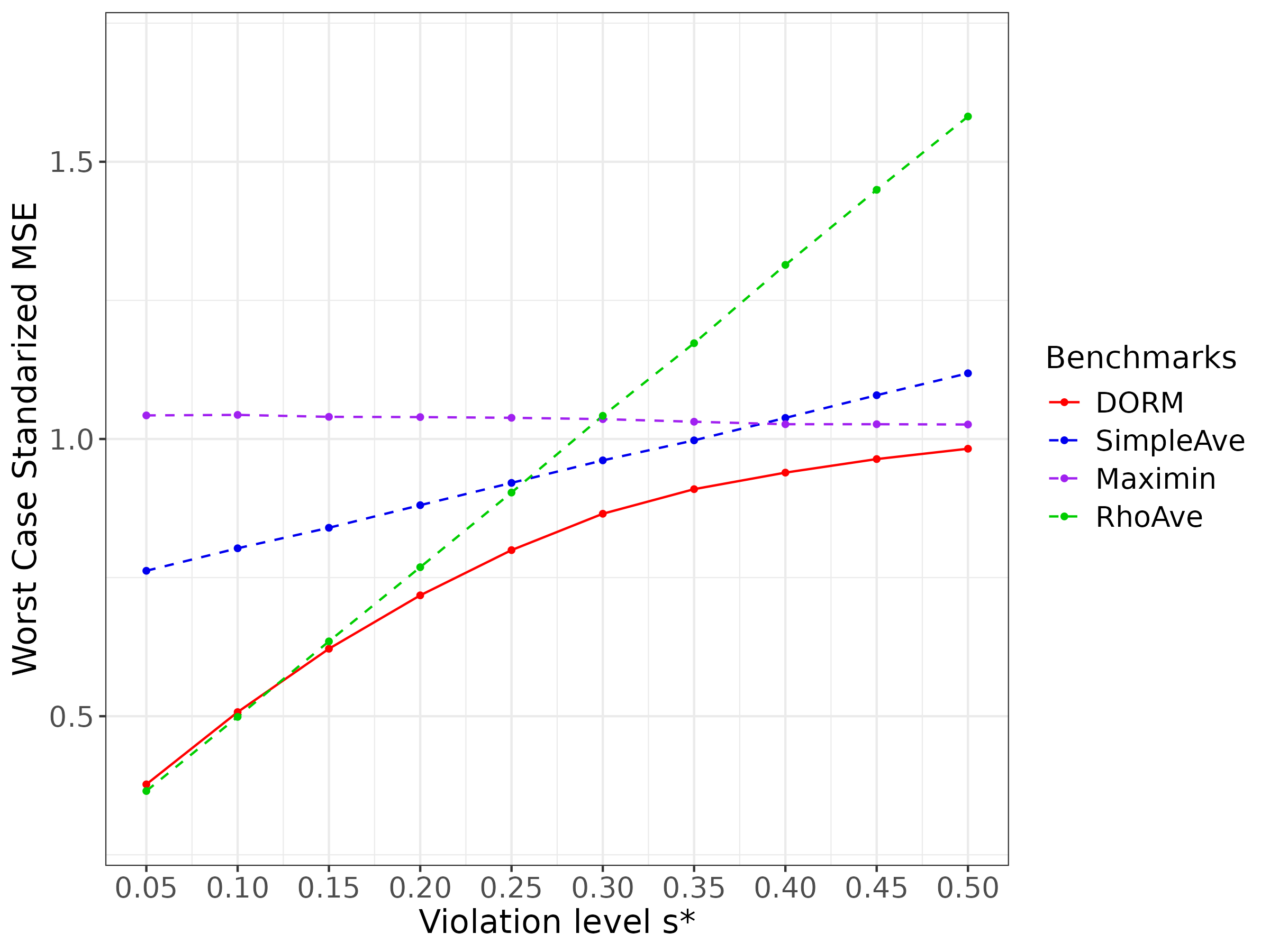}
  \label{fig:L10}

\caption{The worst case performance with numbers of sites $L=10$. The prior probability is $\rho^* = (1/4,0,1/4,0,0,1/4,0,1/4,0,0)$. The red solid line is DORM with tuned $\widehat{\smax}$, and four dashed lines are benchmarks.}
\label{fig:diffL}
\end{figure}

In the fourth scenario \ref{fig:mix3}, we assign equal weights to all five source sites. This is not common in the real-world study, especially when the number of sites $L$ is large. For instance, we do not expect that a mixed blood subject is a mixture of all the potential races (with nonnegligible positive weights); instead we expect that it is a mixture of only a few possible races. In our simulation setting, an equal average of all source sites will lead to a regression coefficient near zero, and more violation of ideal mixture actually will bring the coefficient closer to the origin point. Therefore, both adversarial regression methods will perform better when $s^*$ increases. When $s^* = 0.50$, standardized MSE of DORM is 0.9377, which is about 23\% better than RhoAve and SimpleAve. In Figure \ref{fig:diffL}, we conducted the simulation with $L=10$ different source sites, where the last five sites have moderately different distribution from the first five sites. The performances of the methods are quite similar as the $L=5$ scenario in the Figure \ref{fig:mix0}. 

In Appendix, we conduct an additional simulation study to demonstrate that DORM is less sensitive to weak overlap on the distribution of $X$ between the source and target compared to the common importance weighting strategies for covariate shift adjustment, which is a by-product of our proposed construction. This is because our method utilizes the posterior weight $\eta_l (X;\rho)$ that can actually alleviate the impact of those source samples with their observed covariates deviating far from the target distribution.

\subsection{Tuning of $\smax$ and its sensitivity analysis}
\label{subsec:simu:diffsmax}

\begin{figure}[htbp]
    \centering
\begin{minipage}[c]{0.45\textwidth}
  \includegraphics[width=\textwidth]{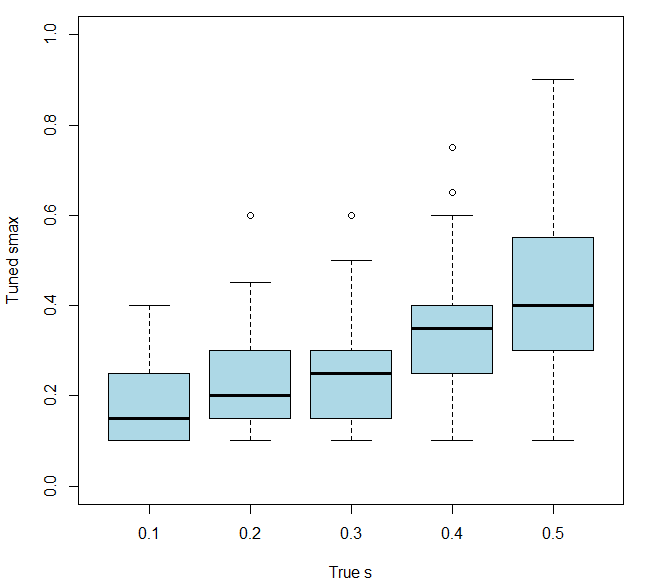}
  \subcaption{$\smax$ tuned by small labeled samples.}
  \label{fig:tuning-smax}
\end{minipage}
\begin{minipage}[c]{0.54\textwidth}
    \includegraphics[width = \textwidth]{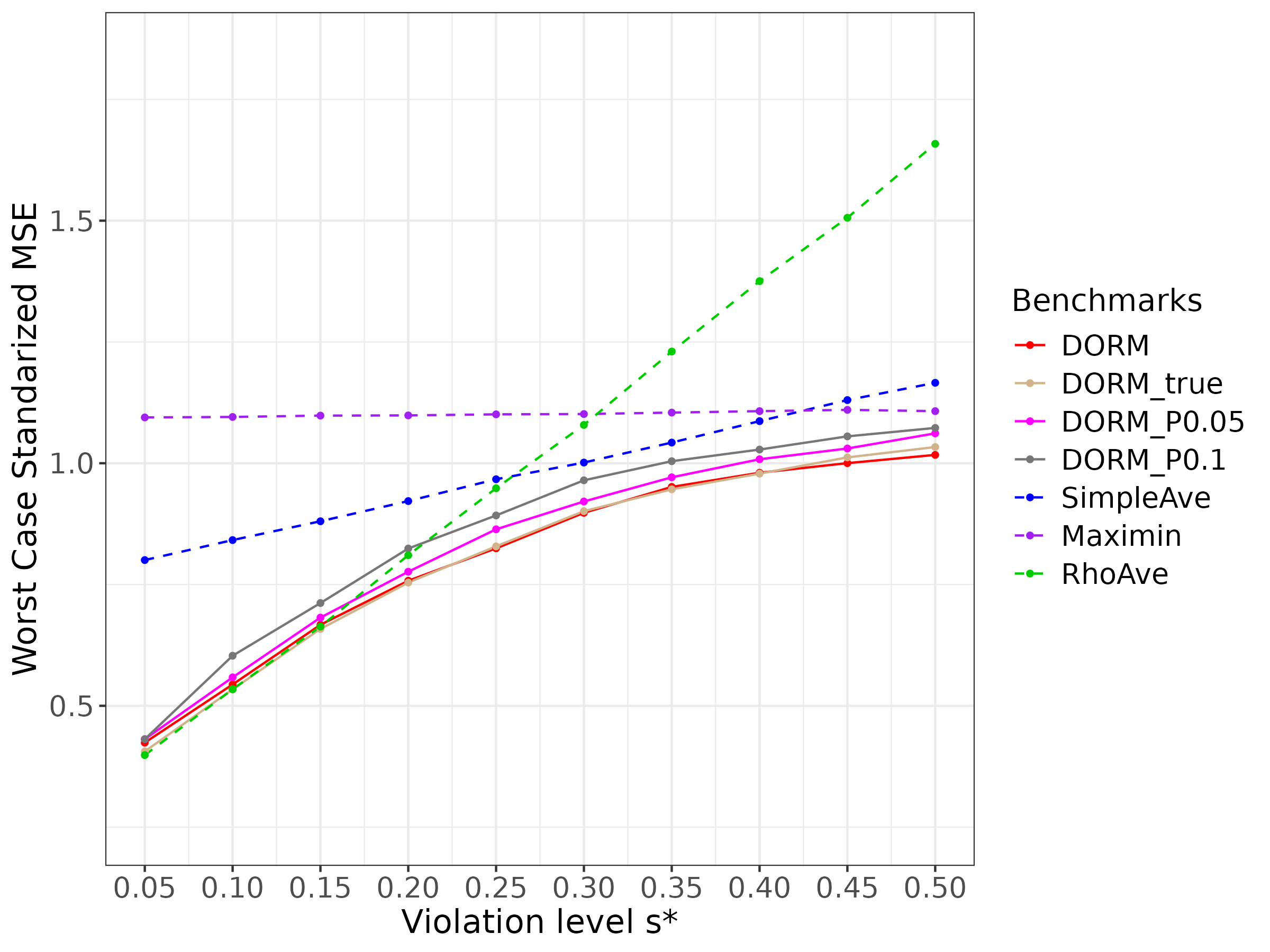}
    \subcaption{Insensitivity of $\smax$}
    \label{fig:main-plot}
\end{minipage}
\caption{(a) shows the boxplot of tuned $\smax$ using $n_\dagger=20$ labeled samples over different true violation levels $s^*$. (b) shows the worst case performance for different methods. Here $L=5, \rho^* = (0.5,0,0.5,0,0)$, which is the same setting as Figure \ref{fig:mix0}. Four solid lines are our methods with different choice of parameter $\smax$, and four dashed lines are the benchmarks. }
\end{figure}

In this subsection, we demonstrate the effectiveness of our tuning strategies for $\smax$ introduced in Section \ref{sec:tuning} and study the sensitivity of our approach to the choice of $\smax$. In our data generation setup, the parameter $s^*$ characterizes the deviation of the target distribution from the joint source-mixing distribution. Figure \ref{fig:tuning-smax} shows the performance of our tuning procedure for $\smax$ with a small number of labeled samples ($n_\dagger = 20$). Under different values of the violation level $s^*$, our tuning procedure can produce $\widehat{\smax}$ relatively close to \(s^*\), demonstrating the validity and effectiveness of tuning. For example when true $s^* = 0.2$, over 80\% of tuned $\widehat{\smax}$ fall into $[0.1,0.3]$. Appendix includes results of tuning with some imperfect surrogate outcomes as described in Section \ref{sec:tuning}.

In Figure \ref{fig:main-plot}, we plot the worst-case standardized MSE of DORM with different ways of choosing or specifying $\smax$ as well as the benchmarks. \texttt{DORM} uses $\widehat{\smax}$ tuned with $n_\dagger = 20$ labels; \texttt{DORM\_true} sets $\smax =s^*$; \texttt{DORM\_P0.05} represents the largest error with $\smax$ in a range $\pm$ 0.05 of the true $s^*$, and \texttt{DORM\_P0.1} represents the largest error with $\smax$ in a range $\pm$ 0.1 of the true $s^*$ value. We observe the difference between the four solid lines is small, with a discrepancy of less than 8\%. \texttt{DORM} is almost identical to the line \texttt{DORM\_true}, which demonstrates the effectiveness of our tuning strategy, even with a very small set of true labels. \texttt{DORM\_P0.05} and \texttt{DORM\_P0.1} indicate that a moderate deviation from $s^*$ will not substantially affect the performance of the DORM and our method is not sensitive to the selection of $\smax$. 

\subsection{High-dimensional domain adaptation}
\label{subsec:simu:highd}

In this subsection, we will compare our method DORM with state-of-the-art domain adaptation algorithms: TransLasso \citep{li2022transfer}, TransGLM \citep{tian2023transfer} and Profile Transfer Learning \citep{lin2024profiled} as summarized in Section \ref{simu:benchmarks}. It is important to note that DORM does not need any labels from the target domain in its training procedure except for tuning $\smax$, while all above-mentioned methods require a decent amount of target labels. We will compare our method with these benchmarks under different setups of the labeled sample size on the target domain. Recall that DORM uses the labeled samples only for the tuning of $\smax$. 

\begin{figure}[htbp]
    \centering
\begin{minipage}[c]{0.47\textwidth}
  \includegraphics[width=\textwidth]{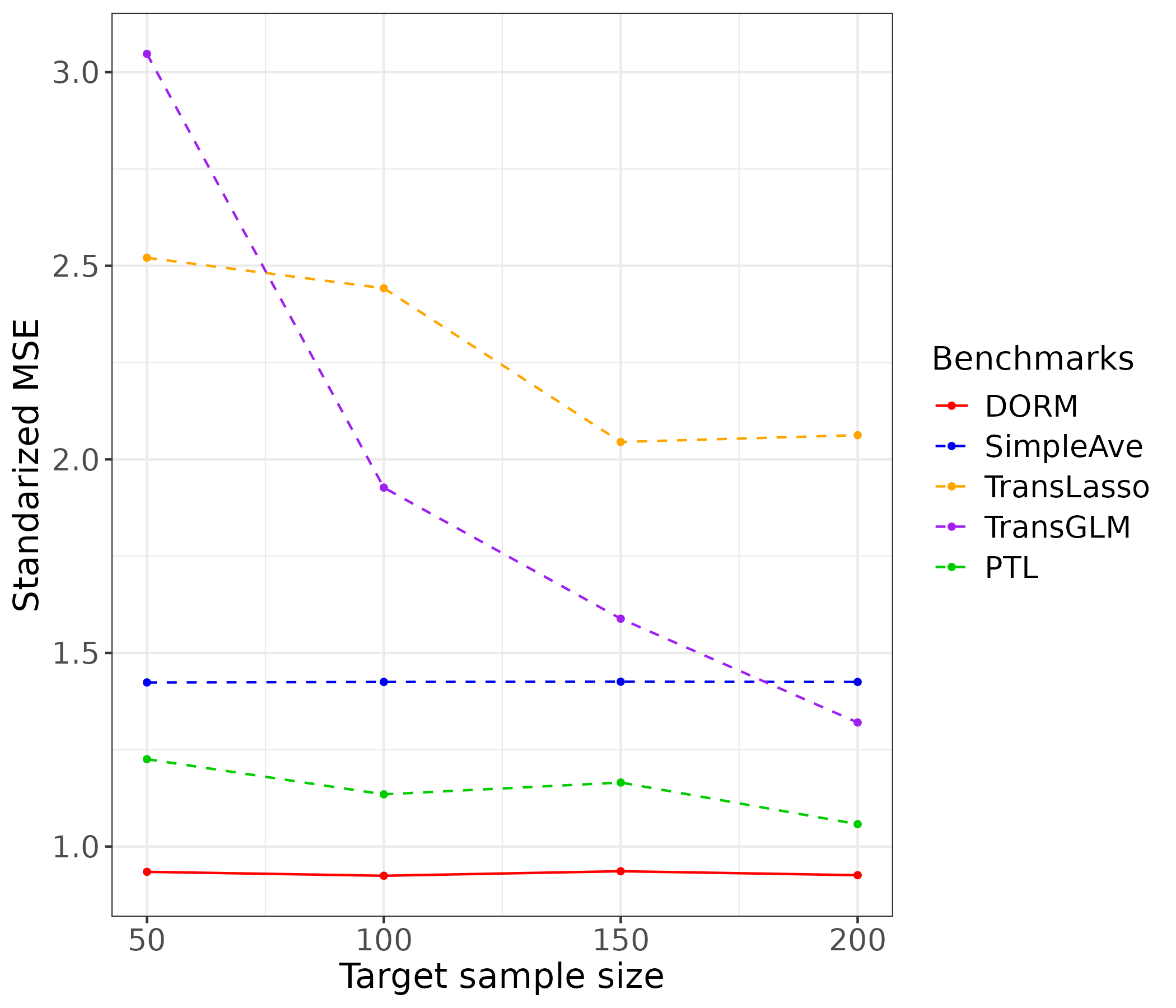}
  \subcaption{Target data evaluation}
  \label{fig:highdim1}
\end{minipage}
\begin{minipage}[c]{0.47\textwidth}
  \includegraphics[width=\textwidth]{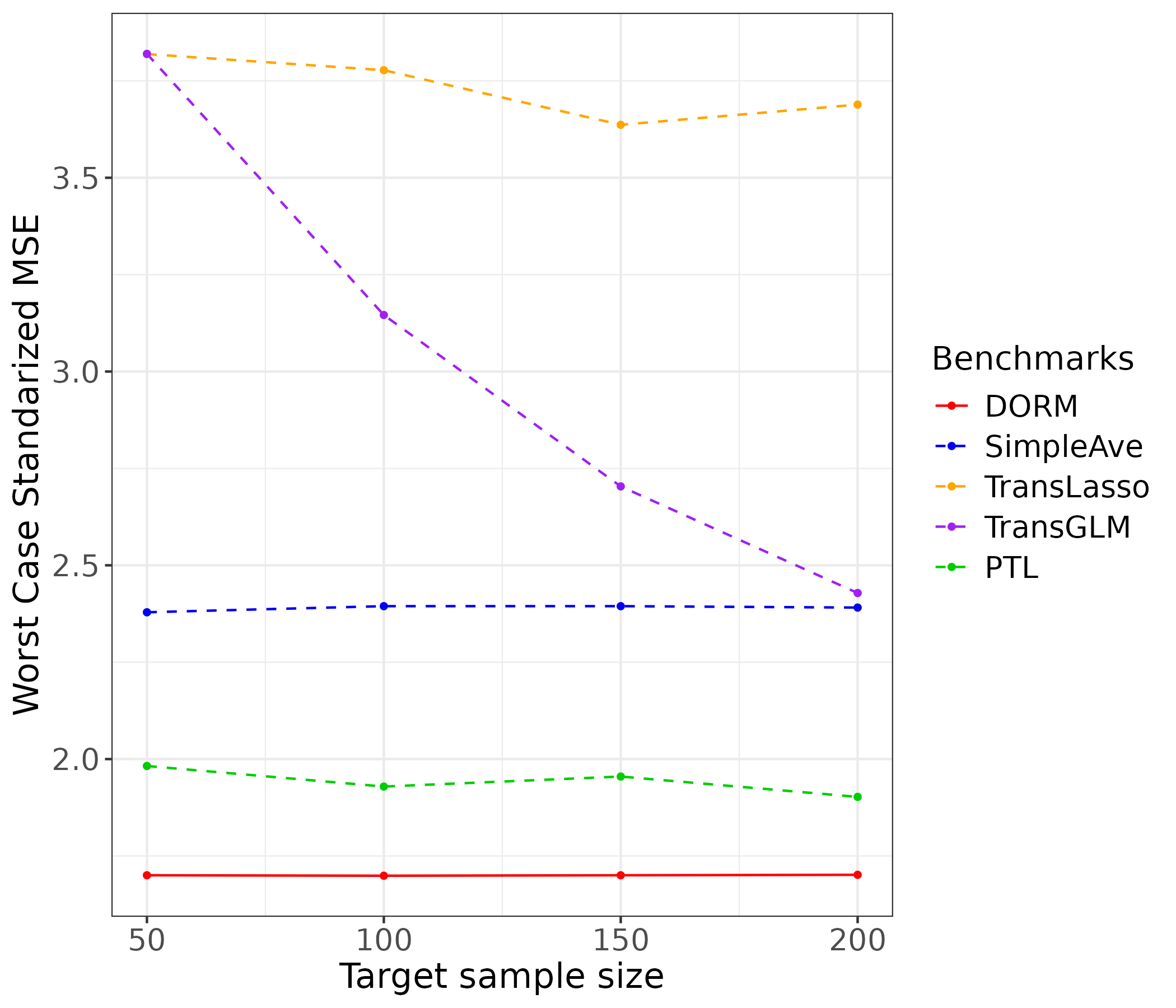}
  \subcaption{Worst case performance}
  \label{fig:highdim2}
\end{minipage}
\caption{Performance of DORM and other high-dimensional domain adaptation methods on the target data. $x$-axis is the number of samples on the target site. For (a), $y$-axis is the standardized MSE evaluated on the target data. For (b), $y$-axis is the worst case standardized MSE on the generated test data over 100 independent trials.}
\label{fig:highdim}
\end{figure}

We focus on a high-dimensional regression setting and set $A = X$ so the dimension of $A$ is $q = 200$. In this setting, we implement DORM with the lasso regularization as described in Appendix. The source mixture weight is chosen as $\rho^* = (0.5,0,0.5,0,0)$, and the violation level $s^* = 0.1$. In Figure \ref{fig:highdim}, we present standardized MSE evaluated on the target data as well as the worst case standardized MSE over different $\delta^*$. On both metrics, DORM performs better than the other domain adaptation methods. For example, even with the largest sample size $n_0 = 200$, DORM achieves a standardized MSE of 0.9260 evaluated on the target, while TransLasso's is 122.67\% higher than DORM, TransGLM is 42.58\% higher, and PTL is 15.30\% higher. The worst-case performance metric gives a similar result, indicating the effectiveness and robustness of DORM. 

The MSEs of TransLasso, TransGLM, and PTL all increase as the target labeled sample size $n_0$ decreases, while the performance of DORM barely changes with this label size. This is because we only use the target labels for tuning of $\smax$ and this low-dimensional procedure is much less sensitive to the reduction of $n_0$ compared to TransLasso and others. This point has been demonstrated in Section \ref{subsec:simu:diffsmax}. For SimpleAve, $\qlh$ is not influenced by $Y\uz_i$, so its simple aggregation is also constant over different target samples. In our setup. PTL performs better than TransLasso and TransGLM mainly because the latter two are based on the vanishing difference of the regression coefficients between the target and a certain subset of sources for knowledge transfer, overlooking the source-mixing structure of target distribution. PTL regresses the target response on the transferred feature, leading to the profiled responses and calculating the residual. However, PTL still shows a worse performance than DORM since it is ``misspecified'' in our joint-source-mixing scenario, under which the target model coefficient vector is not necessarily a linear combination of the source model coefficients.

\begin{figure}[htbp]
    \centering
\begin{minipage}[c]{0.47\textwidth}
  \includegraphics[width=\textwidth]{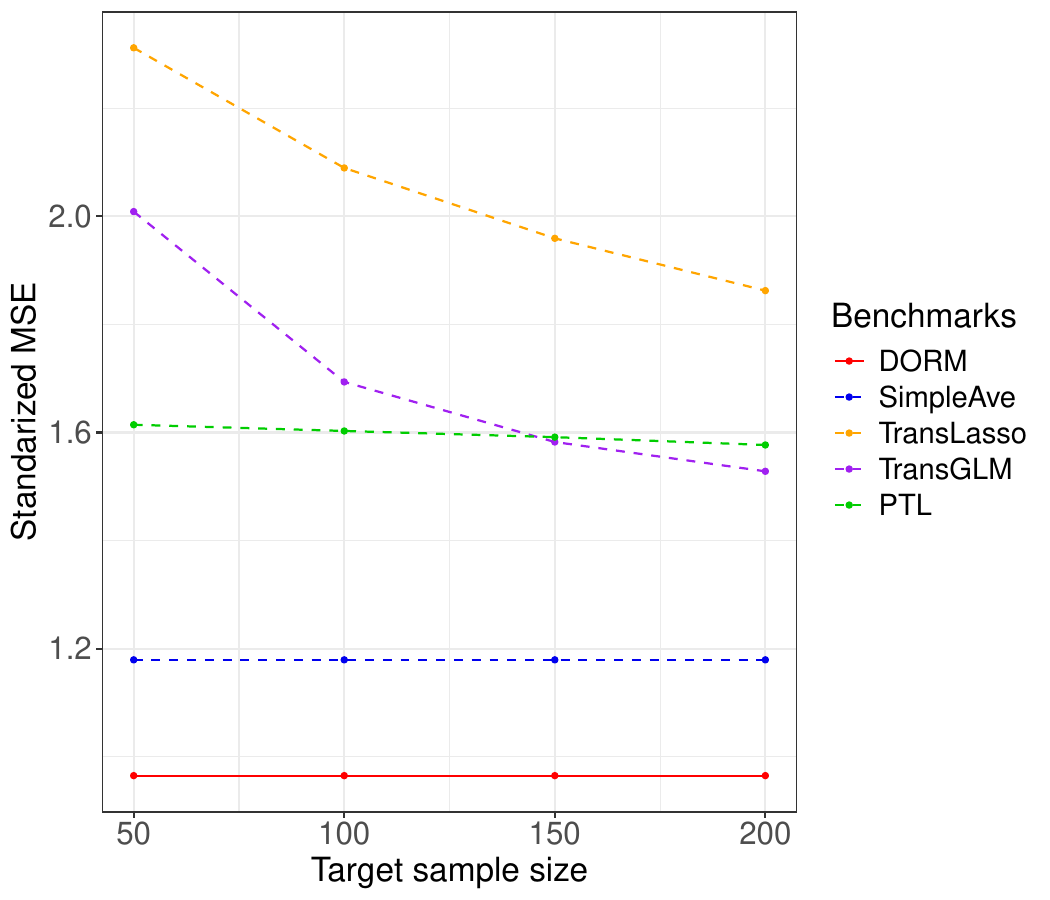}
  \subcaption{Target data evaluation}
\end{minipage}
\begin{minipage}[c]{0.47\textwidth}
  \includegraphics[width=\textwidth]{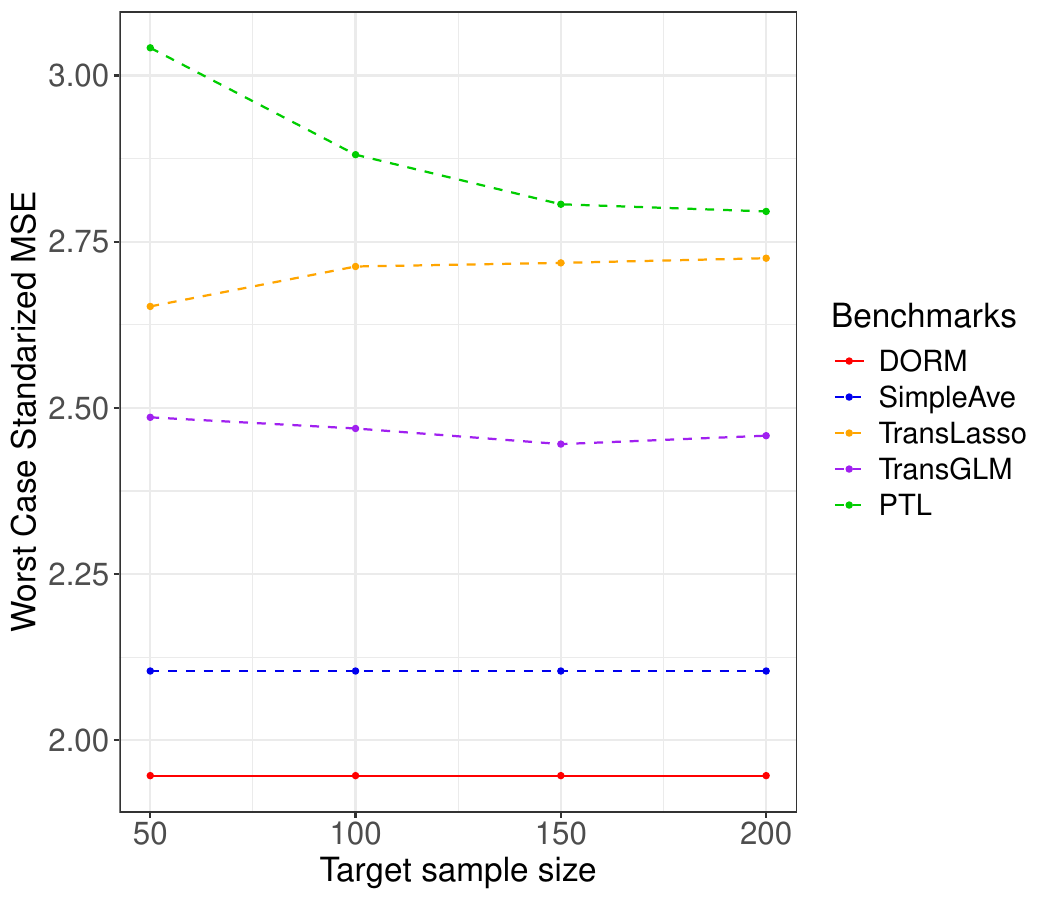}
  \subcaption{Worst Case performance}
\end{minipage}
\caption{Performance of DORM and existing methods under contamination data generation model (\ref{eq:contam1}). $x$-axis is the number of samples on the target site. For (a), $y$-axis is the standardized MSE evaluated on the target data. For (b), $y$-axis is the worst case standardized MSE on the generated test data over 100 independently generated random contamination distribution.}
\label{fig:contam}
\end{figure}

We also evaluate the predictive performance under the contaminated data generation model as indicated by (\ref{eq:contam1}). The contamination distribution $\PP_{Y|X}^{(\varepsilon)}$ is set different from any linear combination of source conditional distributions $\PP_{Y|X}\ul$, and the detailed setup for the target data generation can be found in Appendix. We retain the prior weight as $\rho^* = (0.5,0,0.5,0,0)$, with the true violation level $s^* = 0.1$. Figure \ref{fig:contam} presents the standardized MSE evaluated on the target data and the worst-case standardized MSE. On both evaluation metrics, DORM outperforms other benchmarks. With the largest target labeled sample size of $n_0 = 200$, DORM achieves a standardized MSE of 0.9643, while TransLasso’s is 93.75\% higher, TransGLM’s is 58.34\% higher, and PTL’s is 63.54\% higher. The worst-case performance shows similar trends, demonstrating DORM’s robustness to contamination of the source mixture on the target. 


\section{Real Data Analysis}\label{sec:realdata}
\subsection{Background}

We validate our proposed DORM approach using the high-density lipoprotein (HDL) lab test data from Mass General Brigham (MGB) and UK Biobank (UKB) along with the genetic information. It is believed that the genetic underpinnings of mean lipoprotein diameter differ by race/ethnicity. \cite{frazier2013genetic} found that variation across the intronic region of the LIPC gene was suggestively associated with mean HDL diameters only in Caucasians. In our analysis, among the 195 SNPs reported to be associated with mean HDL diameter in \cite{frazier2013genetic}, we focus on those with minor allele frequency larger than 0.1 in the UKB and MGB cohorts, which ends up with 27 SNPs. We then build a linear model on fasting mean HDL diameters for each race-gender subgroup, adjusted for age. The sample size for each subgroup in MGB and UKB is presented in Table \ref{tab: sample size}. In each cohort, other than the six main racial groups serving as the source sites, there are a small number of mixed-race groups (e.g., White-Asian and White-Black in UKB). Considering these mixtures of sources as the target populations, it is reasonable to assume that the target model is equal/close to the mixture of source models built on the main racial groups. For each target, we randomly sample 50 data points to form the training target label (used to tune DORM and train other benchmarks) and treat the rest as the validation data set to evaluate the prediction performance on the target site. 

\begin{table}[H]
\caption{Sample size of race-gender subgroups in UKB and MGB data.}
\begin{subtable}{1\linewidth}
\centering
  \caption{Sample size of six source sites.}
  \small
    \begin{tabular}{c|cc|cc|cc}
    \toprule
    \multirow{2}[2]{*}{} & \multicolumn{2}{c|}{White} & \multicolumn{2}{c|}{African} & \multicolumn{2}{c}{Asian} \\
          & Female & Male  & Female & Male  & Female & Male \\
    \midrule
    UKB   & 215905 & 184841 & 3753  & 2872  & 4491  & 4949 \\
    MGB   & 17370 & 15155 & 1400  & 810   & 581   & 402 \\
    \bottomrule
    \end{tabular}%
\vspace{0.5cm}
  \centering
  \caption{Sample size of different target sites.}
    \small
    \begin{tabular}{c|cc|cc|cc|cc|cc}
    \toprule
    \multirow{2}[2]{*}{} & \multicolumn{2}{c|}{Other} & \multicolumn{2}{c|}{Unknown} & \multicolumn{2}{c|}{White-Asian} & \multicolumn{2}{c|}{White-Black} & \multicolumn{2}{c}{Mix} \\
          & Female & Male  & Female & Male  & Female & Male  & Female & Male  & Female & Male \\
    \midrule
    UKB   & 2132  & 1670  & 655   & 866   & 402   & 299   & 568   & 307   & 570   & 351 \\
    MGB   & 1026  & 494   & 869   & 611   & 0     & 0     & 0     & 0     & 0     & 0 \\
    \bottomrule
    \end{tabular}%
\end{subtable}%
\label{tab: sample size}
\end{table}

In addition, to evaluate the distributional robustness and generalizability of domain adaptation methods, we choose the same race-gender subgroup as in the training target but from a different data origin to be the validation data set. For example, the training target includes UKB unknown females while the validation population comes from MGB unknown females. Since MGB only contains two types of mixture-race population (i.e., other and unknown), the generalizability analysis focuses on the model portability in different combinations of genders and the two races across two healthcare systems. Ideally, we would expect a distributional-robust model to pertain consistently good performance even if future data come from a shifted population mildly away from the training target. We include DORM, SimpleAve, RhoAve, Maximin, TransLasso, TransGLM and PTL as introduced in Section \ref{simu:benchmarks}, with the coefficient of determination $R^2$ on the testing samples used as the evaluation metric.

\subsection{Results}
\begin{figure}[htbp]
    \centering
    \includegraphics[width=1\textwidth]{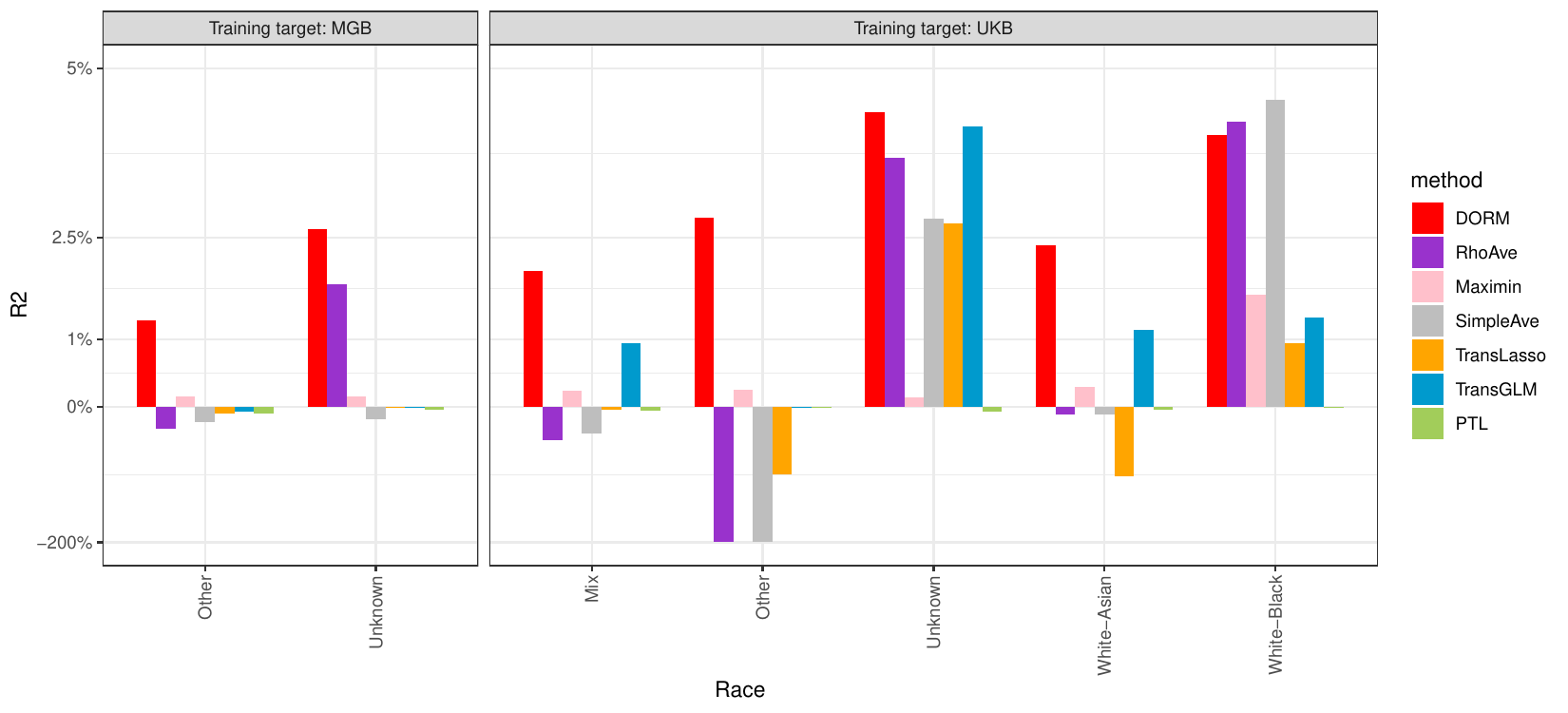}
    \caption{$R^2$ of different methods when both the training target data and the validation data come from the same cohort. Source sites include 6 major race-gender groups in UKB (European female/male, African female/male, Asian female/male). $R^2$ less than $-200\%$ has been truncated to $-200\%$.}
    \label{fig: real data 1}
\end{figure}

In Figure \ref{fig: real data 1}, we present $R^2$ of DORM and other methods when both the training target and the validation data come from the same population, averaged over genders. DORM outperforms other benchmarks with the highest $R^2$ (MGB mean $R^2$: 1.95\%; UKB mean $R^2$: 3.12\%), while SimpleAve and RhoAve suffer from a large negative $R^2$ especially when the training target is UKB Other female/male. Maximin over-shrinks the coefficient to zero, which leads to a small $R^2$ close to zero (MGB mean $R^2$: 0.154\%; UKB mean $R^2$: 0.514\%). The three domain adaptation methods, TransLasso (MGB mean $R^2$: -4.86\%; UKB mean $R^2$: -40.6\%), TransGLM (MGB mean: $-3.94\%$; UKB mean: $1.50\%$), and PTL (MGB mean: $-7.17\%$; UKB mean: $-3.66\%$) fail to train an efficient model due to the limited number of target labels and site heterogeneity, ending up with poor transferability.

\begin{figure}[H]
    \centering
    \includegraphics[width=0.7\textwidth]{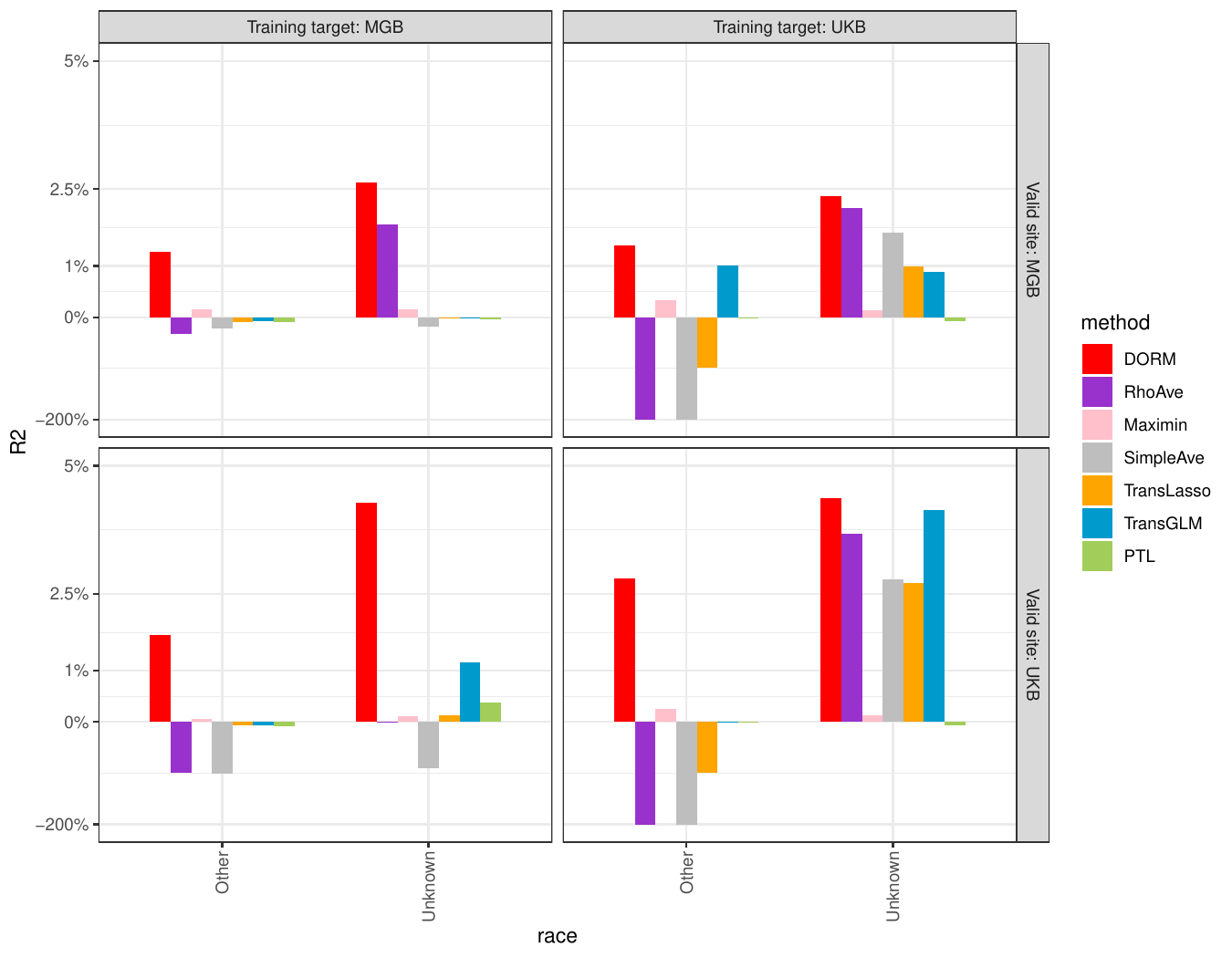}
    \caption{$R^2$ of different methods when the training target data come from a certain race-gender in one site, and the validation data contain data from the same race-gender subgroup in the same site (diagonal panels) or in a different site (off-diagonal panels). $R^2$ is taken average over genders and those with less than $-200\%$ has been truncated to $-200\%$.}
    \label{fig: real data 2}
\end{figure}

Figure \ref{fig: real data 2} further illustrates the generalizable domain adaptation of different models. The off-diagonal panels show the $R^2$ when the training and validation cohorts originate from the same race-gender subgroup but in different healthcare systems. Among all methods, the best performance persists in DORM even when the validation population shifts away from the target population (MGB to UKB: 2.98\%; UKB to MGB: 1.88\%). Maximin (MGB to UKB: 0.08\%; UKB to MGB: 0.24\%) exhibits the second highest positive $R^2$ on average, indicating a certain level of domain adaptation. However, its overall performance for different validation populations is still inferior with $R^2$ close to zero, probably due to the over-conservative design of the uncertainty set. Other than TransGLM (MGB to UKB: -2.38\%; UKB to MGB: 0.95\%), the rest of the benchmark methods suffer from negative $R^2$ in both directions, indicating a lack of generalizable domain adaptation.

\section{Discussion}

We develop DORM, a novel framework for multi-source domain adaptation targeting source-mixture sub-populations such as mixed ethnicity subgroups in our biobank application. DORM takes advantage of the source-mixing structure for effective knowledge transfer under the scarcity or even absence of observations of $Y$ in the target sample. It also incorporates group adversarial learning in order to achieve distributional robustness to the violation of such source-mixing assumptions and improve out-of-distribution generalizability. Meanwhile, our DML approach serves as a novel tool for nuisance-error-robust and efficient estimation in the presence of covariate shift and mixture structure. Importantly, in both the simulation and real-world studies, DORM shows improvement over state-of-the-art multi-source domain adaptation methods in terms of predictive performance and generalizability.

Finally, we discuss several potential future directions of our work. First, it would be useful to accommodate in DORM generalized linear models (GLMs) and general machine learning models for $Y\sim A$. This could be naturally realized by replacing the linear and least squares reward function $R_{\PP}(\beta)$ with those of GLMs and non-parametric prediction models. Nevertheless, new challenges may arise in both optimization and statistical interpretation. Second, for the interval estimation and testing of our population-level model coefficients, one could potentially leverage the resampling inference approach of \cite{guo2022statistical} to address the non-regularity issue of the maximin estimation. Third, our construction strategy of $\eta_l(x)$ may produce inaccurate approximations of $\PP_X\uz$ on target under severe misspecification of the covariate mixing structure (\ref{eq:xmix}). This issue could also not be examined or identified with the data in our current framework. Fixing this problem may require more advanced techniques for robust learning of mixture distributions, which warrants future research.

\bibliographystyle{apalike}
\bibliography{library}

\end{document}